\documentclass[oneside,english,latin9,usenames,table,dvipsnames,svgnames]{aa}
\usepackage[T1]{fontenc}
\usepackage{geometry}
\usepackage{color}
\usepackage{babel}
\usepackage{array}
\usepackage{textcomp}
\usepackage{url}
\usepackage{amstext}
\usepackage{amssymb}
\usepackage{graphicx}
\usepackage[unicode=true,pdfusetitle,
 bookmarks=true,bookmarksnumbered=false,bookmarksopen=false,
 breaklinks=false,pdfborder={0 0 0},pdfborderstyle={},backref=false,colorlinks=true]
 {hyperref}
\hypersetup{
 urlcolor=blue,citecolor=blue}

\makeatletter

\providecommand{\tabularnewline}{\\}
\usepackage{pifont}
\usepackage[usenames,dvipsnames,svgnames,table]{xcolor}
\definecolor{darkred}{rgb}{0.55, 0.0, 0.0}
\definecolor{darkblue}{rgb}{0.03137255087494851, 0.18823529779911066, 0.41960784792900085}
\definecolor{mygray}{HTML}{444444}
\definecolor{darkgreen}{rgb}{0.0, 0.2666666805744171, 0.10588235408067699}
\newcommand\circleone[1][]{{\color{mygray}\ding{172}#1}}
\newcommand\circletwo[1][]{{\color{darkblue}\ding{173}#1}}
\newcommand\circlethree[1]{{\color{orange}\ding{174}#1}}
\newcommand\circlefour[1]{{\color{darkgreen}\ding{175}#1}}
\newcommand\circlefive[1]{{\color{darkred}\ding{176}#1}}

\newcommand\XARS{{\texttt{XARS}}}
\newcommand\xspec{{\texttt{Xspec}}}
\newcommand\sherpa{{\texttt{Sherpa}}}
\newcommand\BXA{{\texttt{BXA}}}
\newcommand\Geant{{\texttt{Geant4}}}

\newcommand\model[1]{{\textsc{#1}}}
\newcommand\uxclumpy{{\model{UxClumpy}}}
\newcommand\SPHERE{{\texttt{SPHERE}}}
\newcommand\BNTORUS{{\texttt{BNTORUS}}}
\newcommand\MYTORUS{{\texttt{MYTORUS}}}

\newcommand{\NH}{ {N_{\rm{H}}} }

\DeclareMathOperator{\paramCircinusnhHiLo}{31_{-16}^{+25} \times {10}^{24}}

\DeclareMathOperator{\paramCircinusphoindex}{1.81 \pm 0.10}
\DeclareMathOperator{\paramCircinusecut}{209 \pm 69}

\DeclareMathOperator{\paramNGCfournhHiLo}{26_{-20}^{+25} \times {10}^{24}}

\DeclareMathOperator{\paramNGCfourphoindex}{2.29 \pm 0.14}

\DeclareMathOperator{\paramESOnh}{(1.6 \pm 0.1) \times {10}^{23}}

\DeclareMathOperator{\paramESOphoindex}{1.84 \pm 0.04}

\DeclareMathOperator{\paramESOctkcover}{28 \pm 7}

\makeatother

\usepackage{listings}

\begin{document}
\title{X-ray spectral and eclipsing model of the clumpy~obscurer in active
galactic nuclei}
\abstract{We present a unification model for a clumpy obscurer in active galactic
nuclei (AGN) and investigate the properties of the resulting X-ray
spectrum. Our model is constructed to reproduce the column density
distribution of the AGN population and cloud eclipse events in terms
of their angular sizes and frequency. We developed and release a generalised
Monte Carlo X-ray radiative transfer code, \texttt{XARS}, to compute
X-ray spectra of obscurer models. The geometry results in strong Compton
scattering, causing soft photons to escape also along Compton-thick
sight lines. This makes our model spectra very similar to the Brightman
\& Nandra \texttt{TORUS} model. However, only if we introduce an additional
Compton-thick reflector near the corona, we achieve good fits to \emph{NuSTAR}
spectra. This additional component in our model can be interpreted
as part of the dust-free broad-line region, an inner wall or rim,
or a warped disk. It cannot be attributed to a simple disk because
the reflector must simultaneously block the line of sight to the corona
and reflect its radiation. We release our model as an \texttt{Xspec}
table model and present corresponding \textsc{CLUMPY} infrared spectra,
paving the way for self-consistent multi-wavelength analyses. }
\author{Johannes Buchner\inst{1,2}\thanks{\protect\href{mailto:mailto:johannes.buchner.acad\%40gmx.com}{johannes.buchner.acad@gmx.com}},
Murray Brightman\inst{3}, Kirpal Nandra\inst{4}, Robert Nikutta\inst{5,1},
Franz E. Bauer\inst{1,6,7}}
\institute{Pontificia Universidad Católica de Chile, Instituto de Astrofísica,
Casilla 306, Santiago 22, Chile\and
Excellence Cluster Universe, Boltzmannstr. 2, D-85748, Garching, Germany\and
Cahill Center for Astrophysics, California Institute of Technology,
1216 East California Boulevard, Pasadena, CA 91125, USA \and
Max Planck Institute for Extraterrestrial Physics, Giessenbachstrasse,
85741 Garching, Germany\and
National Optical Astronomy Observatory, 950 N Cherry Avenue, Tucson,
AZ 85719, USA\and
Millenium Institute of Astrophysics, Vicuña MacKenna 4860, 7820436
Macul, Santiago, Chile\and
Space Science Institute, 4750 Walnut Street, Suite 205, Boulder, Colorado
80301}
\date{-Received date / Accepted date}
\titlerunning{Clumpy unified AGN obscurer model}
\authorrunning{Buchner et al.}
\maketitle

\section{Introduction}

X-ray spectroscopy is a powerful method for characterising the accretion
onto supermassive black holes, even when the accretion disk is hidden
behind thick columns of gas and dust. For the most heavily obscured
Compton-thick active galactic nuclei (CT AGN, $\NH\gtrsim10^{24}{\rm cm}^{-2}$)
inference on the accretion and nature of the obscurer requires a correct
spectral model for the obscurer (the `torus'). To date, the explored
range of X-ray spectral models has been limited to simple geometries,
while observations have shown that the real geometry is more complex.

For instance, strong evidence exists today that the obscurer is clumpy
and has a large covering factor. The former is found through X-ray
eclipse events \citep[e.g.][]{Risaliti2002,Markowitz2014}, while
the latter is evident from the large fractions of obscured AGN found
in flux-limited X-ray surveys \citep[e.g.][]{Treister2004,Brightman2014,Ueda2014,Buchner2015,Aird2015}.
Following the orientation-only version of the unification paradigm
\citep[straw-person model,][]{Antonucci1993}, a single obscuring
structure may explain the diversity of obscured and unobscured AGN
as well as eclipse events through sampling different viewing angles.
However, such a strict unification across all AGN is ruled out because
the accretion luminosity is anti-correlated with the obscured fraction
\citep[e.g.][]{Ueda2003,Hasinger2005,Buchner2015}. In both the infrared
and X-ray regime, many studies have therefore attempted to constrain
the covering factor of individual AGN.

\begin{figure*}
\begin{centering}
\includegraphics[width=1\textwidth]{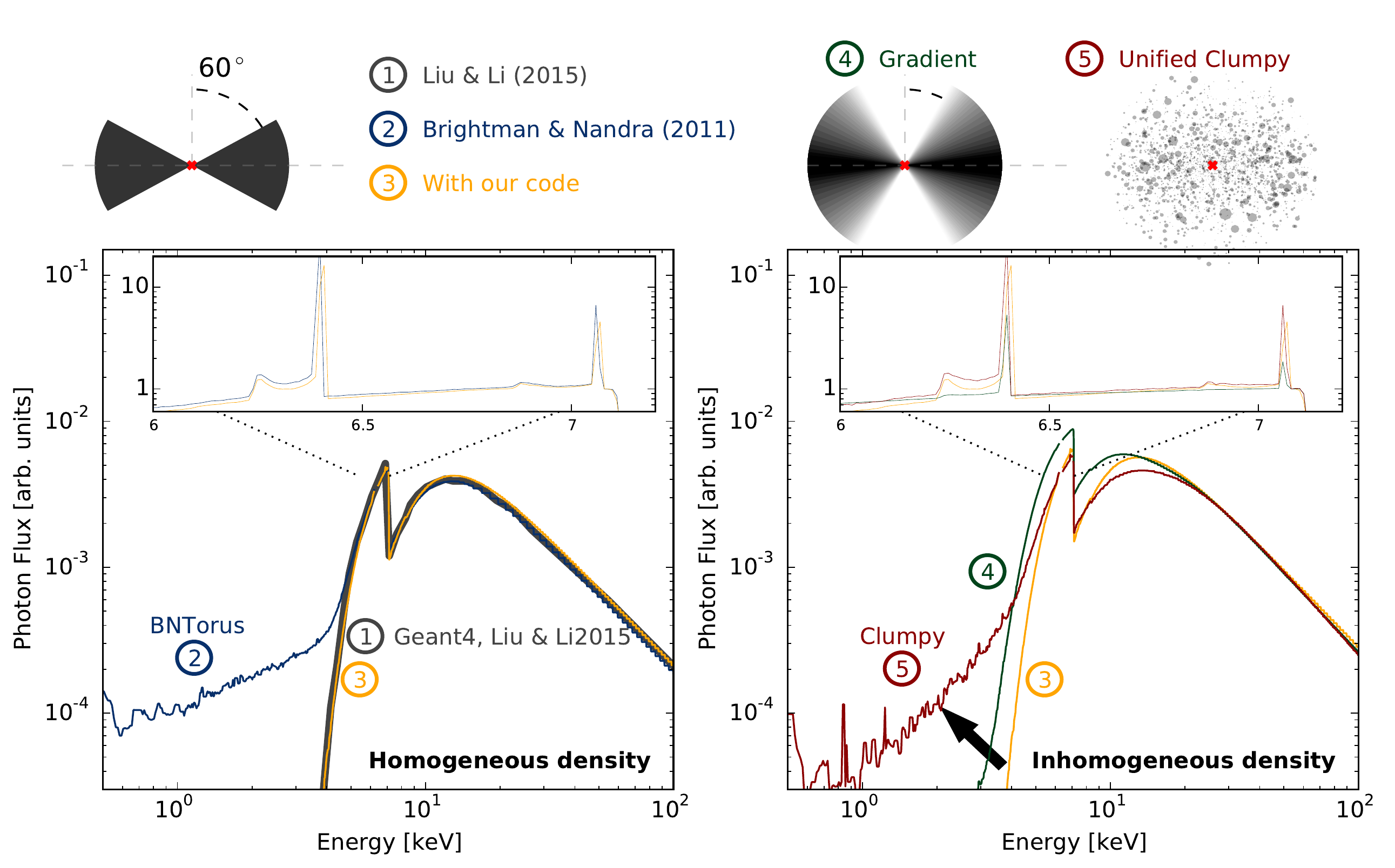}
\par\end{centering}
\caption{\label{fig:spectra}Simulated X-ray spectra, with geometries shown
on top. The \emph{left panel} presents spectra of a Compton-thick
bi-cone cut-out geometry in edge-on view. \citet[shown as \circleone{}]{Liu2015}
demonstrated that the model of \citet[shown as \circletwo]{Brightman2011a}
should not leak soft ($\lesssim4{\rm keV}$) photons. We obtained
with our code \texttt{XARS} (shown as \circlethree{}) the same spectrum
as \citet{Liu2015}. The \emph{right panel} compares this uniform
density toroid with a gradient toroid (shown as \circlefour{}) and
our unified clumpy model (shown as \circlefive{}). The arrow highlights
that, unlike the smooth geometry, leakage of soft photons is possible
in the clumpy torus. In the main plots, we removed spectral lines
to highlight the continuum shape. The \emph{insets} show a zoom around
the Fe K$\alpha$ and $\beta$ features, normalised at $7.07\mathrm{keV}$.
They feature the strongest scattering low-energy tails in the \circlefive{unified
clumpy model}. }
\end{figure*}
Therefore, physically motivated geometric models are needed. Early
modelling efforts \citep[see e.g.][and references therein]{Matt2000a}
of the global, low-resolution X-ray spectra at the time employed simplistic
models \citep[e.g. slab appropriate for cold reflection from accretion disks;][]{Magdziarz1995}.
As knowledge and capabilities in spectral and spatial resolution progressed,
more realistic modelling was developed, allowing refined parameter
constraints and interpretations \citep[e.g.][]{MurphyYaqoobMyTorus2009,Brightman2011a,Ikeda2009}.
Studies of nearby \citep[e.g.][and references therein]{Matt2000}
and distant \citep[e.g.][]{Buchner2014} obscured AGN show that X-ray
spectra are to first order consistent with a photo-electrically absorbed
power law, but additionally require varying degrees of Compton-scattering
to fit the generally flat spectra and the $\sim30\mathrm{keV}$ Compton
hump. This can be seen in the literature, for example by many fits
preferring the model of \citet{Brightman2011a}, \BNTORUS{}, which
produces a flat spectrum for Compton-thick obscured AGN, similar to
that created by cold disk reflection. Some studies even find additional
Compton reflection beyond this model, with examples both in individual
local AGN \citep[e.g.][]{Arevalo2014,Bauer2014} and large X-ray surveys
\citep[e.g.][]{Rivers2013,Buchner2014}. The arrival of \emph{NuSTAR}
\citep{HarrisonNuSTAR2013} revealed, in detail, the complexity of
spectral shapes above $10\mathrm{keV}$ for many local AGN. The analysis
of these data is limited by the currently freely available, easy-to-install
X-ray spectral models of AGN obscurers. The two physically motivated
models most commonly used today are: the \BNTORUS{} model of a bi-conical
sphere cutout, and the `donut' geometry of \citet{MurphyYaqoobMyTorus2009},
\MYTORUS{}. Both assume simple geometries and a smooth distribution
of obscuring material, which cannot explain AGN with time-variable
line-of-sight column densities, commonly interpreted as cloud eclipse
events \citep[e.g.][]{Risaliti2002}. A few clumpy geometry models
have been recently proposed \citep[e.g.][]{Liu2014,Furui2016}, but
have not been used to explain eclipse events. To explore various possible
geometries, interest has grown in the availability of Monte Carlo
codes \citep[e.g.][]{Paltani2017,Balokovic2018}. An extendable open
source Monte Carlo simulation package is not available yet.

To compute X-ray spectra we developed a new Monte Carlo simulation
code, \texttt{XARS}. We begin in §\ref{sec:simcode} by describing
this code and basic results for inhomogeneous media. In §\ref{sec:Model-setup}
we construct a clumpy, unified AGN obscurer model which reproduces
(1) the column density distribution of AGN at various luminosities
and (2) the distribution of cloud eclipse events. Although our initial
model spectra produce strong Compton scattering signatures, in §\ref{sec:Obs}
we demonstrate that we must add another reflection component to fit
\emph{NuSTAR} data. We interpret this result and discuss physical
origins in §\ref{subsec:Interpretation-ring}. Finally, we show how
our clumpy obscurer model allows self-consistent infrared \& high-energy
studies in combination with \textsc{Clumpy} models \citep{Nenkova2008a,Nenkova2008}.

\section{Computation of X-ray spectra in inhomogenous media\label{sec:simcode}}

\subsection{Simulation code XARS}

We compute the emerging X-ray spectrum for several torus geometries
of interest through Monte Carlo simulations. We assume that an isotropically
emitting point-source generates photons at the centre of our obscurer
geometry. Photo-electric absorption, Compton scattering and line fluorescence
(Fe K$\alpha$+K$\beta$; K$\alpha$ of C, O, Ne, Mg, Si, Ar, Ca,
Cr and Ni) are simulated self-consistently as photons pass through
matter. The simulation method has already been described in detail
in \citet{Brightman2011a}, but we provide some computational details
in §\ref{sec:Green-function-computation}. Solar abundances \citep{Anders1989}
are assumed with cross-sections from \citet{Verner1996}.

We make our modular, Python-based simulation code \texttt{\XARS{}}
(X-ray Absorption Re-emission Scattering) publicly available at \url{https://github.com/JohannesBuchner/xars}.
Users can specify their obscurer geometries by implementing a class
which prescribes how photons propagate through the medium, and easily
change the input abundances and emission lines without modifying the
code. Once the Green's functions at each energy grid point and viewing
angle have been computed with adequate signal-to-noise, an input photon
spectrum, for example a powerlaw, can be applied, and an \texttt{\xspec{}}
grid model compiled. 

We verified our code first by comparing with other implementation.
We also verified generated spectra from (1) the \SPHERE{} geometry
of \citet{Brightman2011a} and \citet{Liu2015}, discussed below,
(2) disk reflection off a semi-infinite plane-parallel slab \citep{Magdziarz1995},
discussed in the Appendix, (3) reflection off individual spheres (\citealp{Nandra1994a},
see Appendix~\ref{sec:profiles}) and (4) the biconical cut-out geometry
studied in the next section. As a benchmark, for a Compton-thick sphere,
\texttt{XARS} can compute very high-quality spectra ($10^{9}$ photons
input across 1000 energy bins) in $90$ minutes on a single $2{\rm GHz}$
laptop CPU, and is trivial to parallelise through independent executions.

\subsection{Inhomogenous geometries leak soft photons}

An interesting test case studied in previous works is a sphere with
a biconical cut-out, illustrated in the top left of Figure~\ref{fig:spectra}.
The left panel of Figure~\ref{fig:spectra} compares \circlethree{}
our results with those of the \Geant{}-based code of \citet{Liu2015}
(\circleone{}, grey thick line). For the spectra shown in Figure~\ref{fig:spectra}
we have chosen an edge-on line of sight with column density $\NH=10^{24}{\rm cm}^{-2}$
and a photon index of $\Gamma=1.8$, consistent with \citet{Liu2015}.
We find consistent results. In \citet{Brightman2011a}, the geometry
code erroneously\footnote{J. Buchner \& M. Brightman, priv. comm.}
let photons back-scattered from the inner wall escape without considering
re-entry into the torus \circletwo{}. This affects the geometric
interpretability of fits with this model, and potentially the parameters
inferred from jointly fit components. Nevertheless, it is very interesting
that across the literature this model is found to describe observational
data well \citep[e.g.,][]{Brightman2011a,Brightman2011b,Rivers2013,Buchner2014,Ricci2017a}.

Next we simulated the spectrum of a clumpy model (similar to \textsc{Clumpy}
geometries, see next section for details). We assume spheres of constant
density\footnote{See Appendix~\ref{sec:profiles} for the impact of assuming different
density profiles.}. Our cloud geometry code uses \texttt{LightRayRider} \citep[see][for details]{Buchner2017}
with parallelised and optimised C routines to quickly solve the individual
photon paths through millions of clouds exactly. However, our tests
indicate that beyond $10^{4}$ clouds, the resulting spectrum is not
affected any more by the number of clouds. We generated $\sim10^{9}$
photons for each geometry. For the photons not absorbed inside the
obscurer and escaping to infinity we record energy and direction.
The LOS column density $\NH$ of that direction from the central source
is then computed, and the photon assigned to that particular $\NH$
and altitude angle bin\footnote{The models considered here are azimuthally symmetric.}.
In the right panel of Figure~\ref{fig:spectra} we plot the energy
distribution of photons emerging from directions with LOS column densities
of $\NH=10^{24}{\rm cm}^{-2}$. Photons originally on paths of lower
LOS column densities can be scattered into the LOS. This leads to
a substantial component of soft photons proportional to the scattering
surface (covering fraction) in \circlefive{} clumpy torus models,
as noted in previous works \citep{Liu2014,Furui2016}. We also consider
a smooth torus with gradually varying density layers \circlefour{}
reproducing the same column density distribution. In contrast to the
clumpy model, this geometry blocks soft photons from escaping. In
the clumpy model, fluorescent lines are also scattered farther and
emerge with strong horns in the spectrum (inset plot of Figure~\ref{fig:spectra}).

\section{Clumpy Torus Geometry\label{sec:Model-setup}}

\subsection{Cloud population geometry from obscured fractions}

\begin{figure}
\begin{centering}
\includegraphics[width=1\columnwidth]{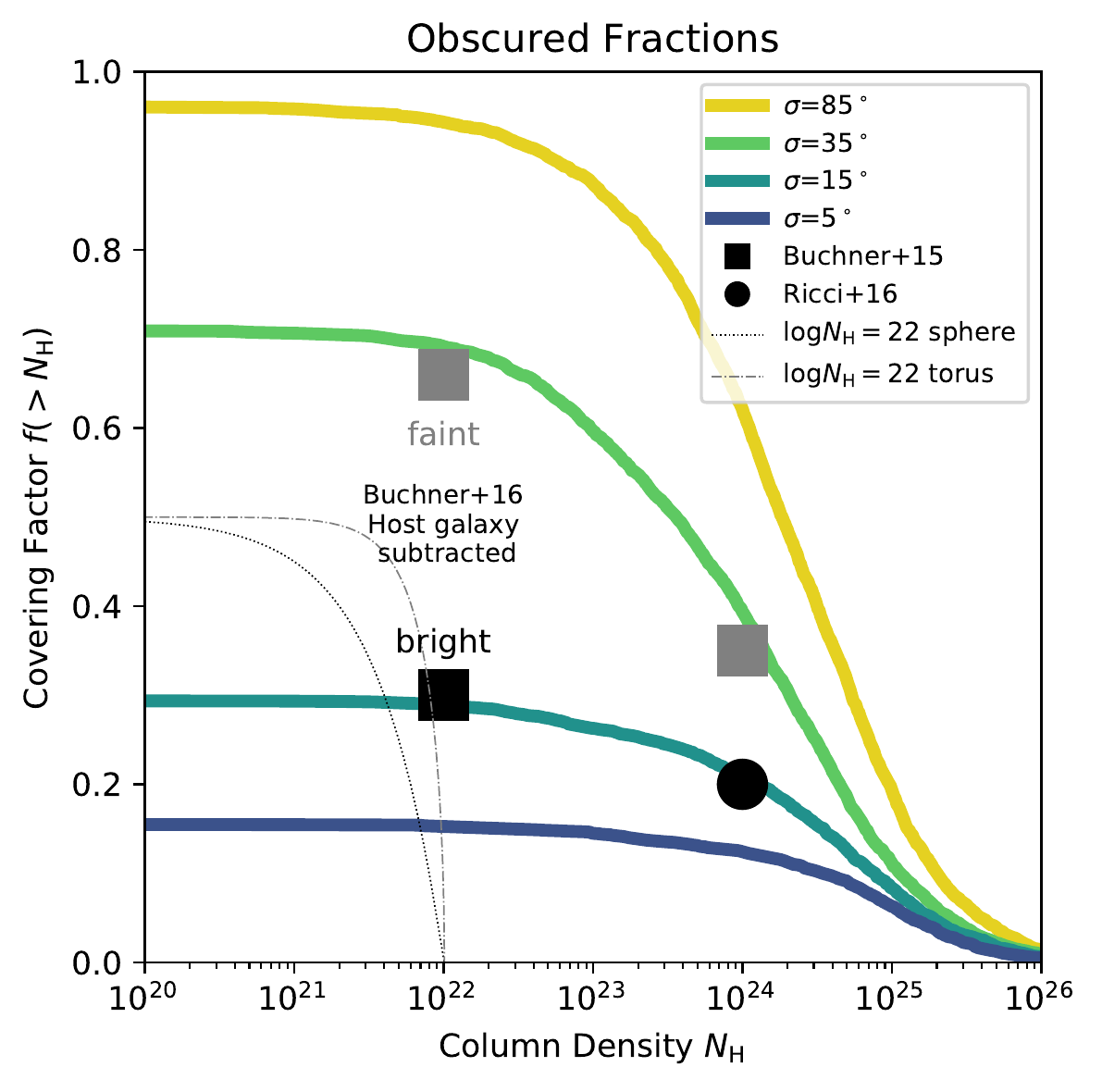}
\par\end{centering}
\caption{\label{fig:Column-density-profile}Column density distribution and
associated covering fractions produced by our clumpy model. The curves
represent variations of the angular width parameter $\sigma$. Obscured
fractions from AGN surveys are shown as black and grey symbols, for
bright and faint AGN, respectively. Varying $\sigma$ allows a single
instance of our model to reproduce the column density distributions
of the population at various luminosities. Thin grey curves show the
comparatively narrow column density distributions of a single sphere
and a torus, each with a maximum crossing column density of $\NH=10^{22}\mathrm{cm}^{-2}$.}
\end{figure}
\begin{figure}
\begin{centering}
\includegraphics[width=1\columnwidth]{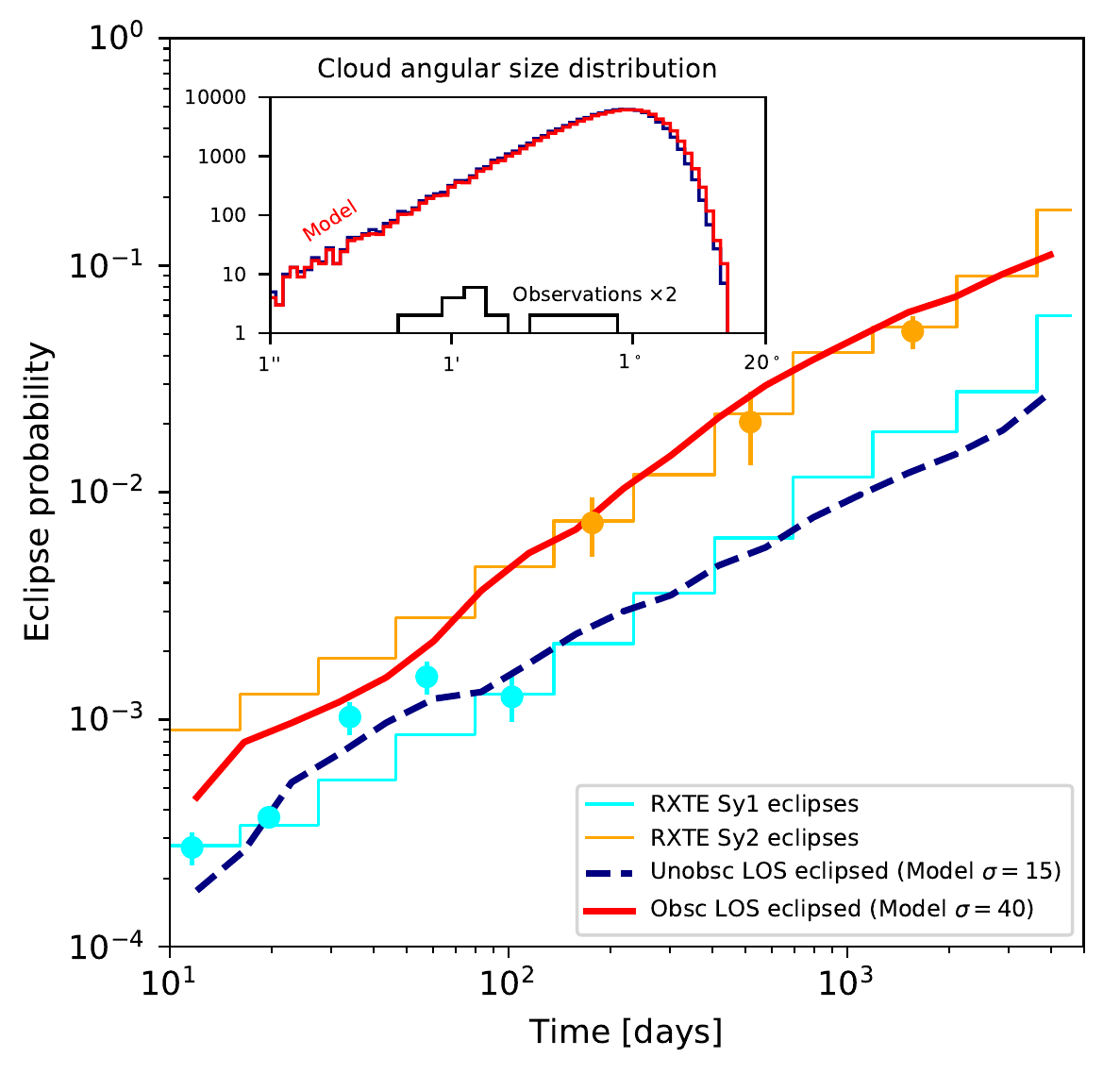}
\par\end{centering}
\caption{\label{fig:eclipses}Predicted rates of eclipses by our model. The
probability for an eclipse in a sightline after a given delay is computed,
with model clouds on random, circular Keplerian orbits. The systematic
study of eclipse event rates measured with the RXTE satellite by \citet{Markowitz2014}
are drawn with orange/cyan error bars (type-1/type-2, respectively),
with steps approximating their observation window function. We analyse
two cases: How frequently an obscured sightline becomes more obscured
(red curve) in a high-covering geometry, and how frequently an unobscured
sightline becomes obscured in a low-covering geometry (dashed blue
curve). The eclipse probability increases with the covering factor.
The angular diameter distribution of the model clouds is shown in
the inset (blue: low-covering model, red: high-covering model). Sizes
of individual observed eclipsing clouds are shown as a black histogram
(Nikutta et al., in prep), covering the low-end tail of the intrinsic
distribution.}
\end{figure}

In principle, the construction of a clumpy obscurer model has infinite
degrees of freedom, as each location can have an arbitrary density.
To simplify this problem, we assume symmetry around one axis and that
the obscurer column density is, at least on average, a monotonic function
with decreasing column density towards the poles. We use the \citet[N08 hereafter]{Nenkova2008}
formalism to describe the distribution and properties of clouds. In
that formalism, the number of clouds ${\cal N}$ seen along a radial
line-of-sight is axi-symmetric, and decreases from ${\cal N}_{0}$
at the equatorial plane with inclination angle $\beta$ towards the
poles, according to a Gaussian function:

\begin{equation}
{\cal N}={\cal N}_{0}\cdot\exp\left\{ -\left(\frac{\beta}{\sigma}\right)^{m}\right\} \label{eq:nenkova-N}
\end{equation}
The angular width, $\sigma$, controlling the torus scale height and
the average number of clouds in the equatorial plane, ${\cal N}_{0}$,
are free parameters a priori. We follow previous studies (e.g. N08)
which adopted a Gaussian profile ($m=2$). Using an exponential distribution
instead does not affect our conclusions.

\begin{figure*}[t]
\includegraphics[viewport=0bp 0bp 360bp 360bp,width=1\columnwidth]{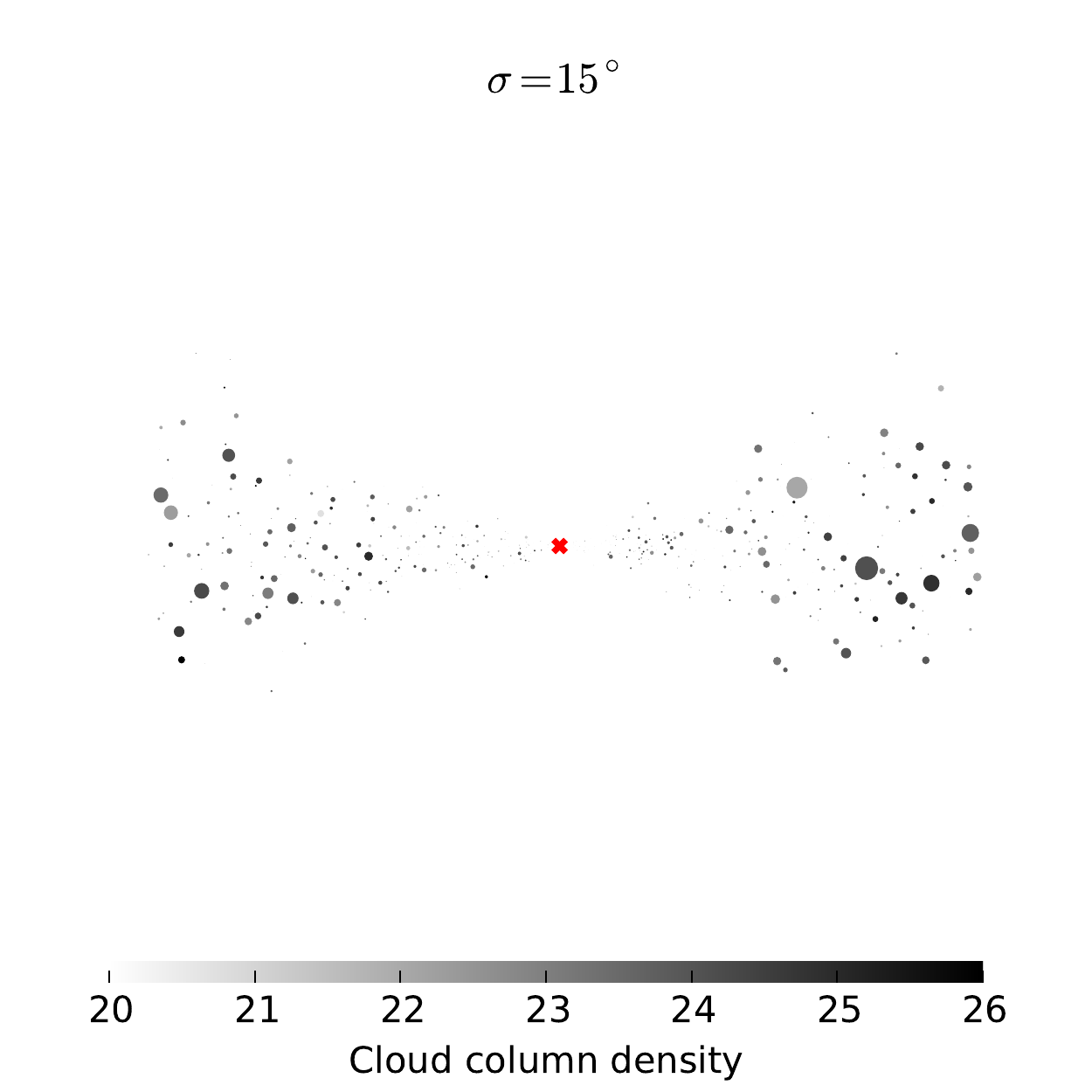}\includegraphics[width=1\columnwidth]{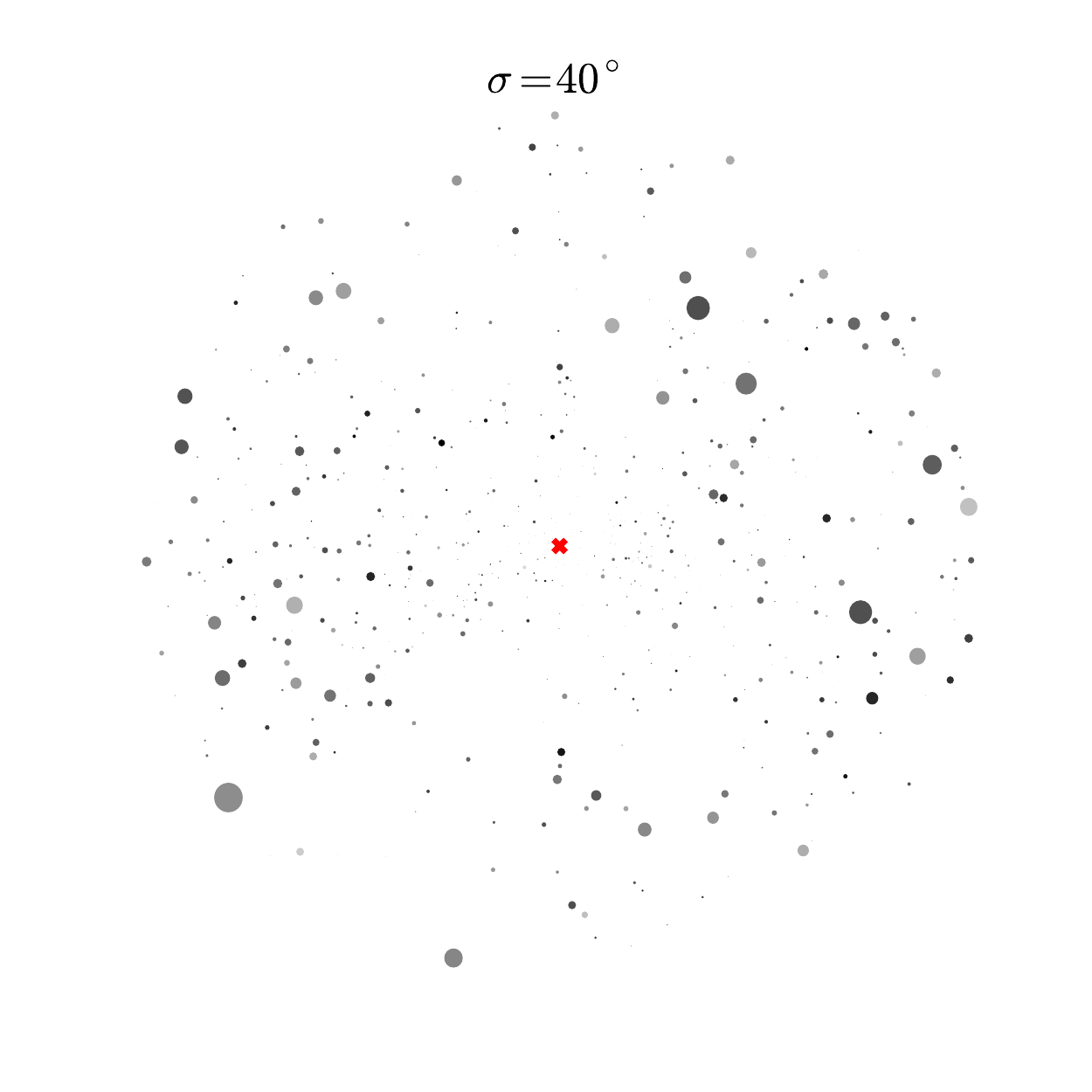}

\caption{\label{fig:Crosssections}Cross-sections of clumpy torus model. The
left and right geometries differ in the vertical extent $\sigma$,
which controls the Gaussian distribution away from the equatorial
plane. The red cross indicates the central point source. The clouds
vary in size and column density (darkness). A fly-through video is
available at \protect\href{https://doi.org/10.5281/zenodo.602282}{https://doi.org/10.5281/zenodo.602282}.}
\end{figure*}
To narrow down the possible geometries in a realistic way, we first
attempt to fit the model to several known constraints. The first constraint
applied is the \textbf{column density distribution}. For luminous
(moderately luminous) AGN, we should reproduce the fractions of $25\%$
($35\%$) for $N_{{\rm H}}\gtrsim10^{24}{\rm cm}^{-2}$ and $30\%$
($65\%$) at $N_{{\rm H}}>10^{22}{\rm cm}^{-2}$ \citep[from][]{Buchner2015,Aird2015,Ricci2016}.
These are intrinsic fractions (observed, but corrected for selection
bias) with the obscuration of the host galaxy subtracted \citep[see][for details]{Buchner2017a}.
Simultaneously, the obscurer should allow, albeit rarely, column densities
up to $N_{{\rm H}}=10^{26}{\rm cm}^{-2}$, since sources with $N_{{\rm H}}>10^{25}{\rm cm}^{-2}$
are known \citep[e.g. NGC 1068,][]{Matt2000a}. However, when assuming
each cloud has the same column density, the column density distribution
produced by Equation~\ref{eq:nenkova-N} is too narrow.

Widening the model column density distribution can be achieved by
allowing at a given luminosity a diversity of opening angles or by
giving the clouds a diversity in column densities. We choose the latter
option, as we prefer to remain with a unification model. Additionally,
absorber variability studies show a wide range of preferentially Compton-thin
column density changes \citep[e.g.][]{Risaliti2002,Risaliti2005,Markowitz2014},
and Compton-thick variations have also been seen when telescopes with
good sensitivity and angular resolution at $>10\mathrm{keV}$ became
available \citep[e.g.,][]{Ricci2016a,Marinucci2016}. We therefore
allow a range of values for the column density of individual clouds,
$N_{{\rm H}}^{{\rm cloud}}$. Clouds are placed randomly according
to the distributions (and re-drawn when overlapping). Fitting the
angular width $\sigma$, the total number of clouds and the parameters
of a log-normal distribution simultaneously, we find a suitable fit
with $10^{5}$ clouds with a 1~dex wide distribution centred at $10^{23}{\rm cm}^{-2}$.
This is shown in Figure \ref{fig:Column-density-profile}. Varying
the angular width $\sigma$ makes our model (coloured thick curves)
reproduce the column density distribution of the bright or faint AGN
population (black and grey points, respectively).

The \textbf{radial cloud distribution} was assumed to be uniform across
two orders of magnitude. While the model is scale-free, we can define
the ratio of outermost-to-innermost radial distances of clouds, $Y=R_{\mathrm{out}}/R_{\mathrm{in}}$.
Observations \citep[e.g.][]{Lira2013,Fuller2016,Garcia-Burillo2016,Ichikawa2016}
indicate that the clouds relevant for the infrared radiation extend
out up to approximately one order of magnitude farther than the dust
sublimation radius, that is, $Y=R_{\mathrm{out}}/R_{\mathrm{d}}\approx10-20$.
For X-rays however, inner ionised clouds\footnote{e.g. those in the broad-line region (BLR)}
can also act as absorbers. We choose $Y=100$ for the X-ray model
computation, so that the X-ray obscurer spans two radial orders of
magnitude, while the UV/optical, dusty obscurer spans only the last
magnitude in radial scale (motivated by \citealp{Tristram2011} and
\citealp{Markowitz2014}). For a consistent infrared model, each cloud
outside the sublimation radius can be assigned an optical depth of
$\tau_{V}$, related to the column density via $\tau_{V}=N_{{\rm H}}^{{\rm cloud}}/\left(2\cdot10^{21}{\rm cm}^{-2}\right)$,
the galactic relationship (\citealp{Predehl1995,Nowak2012}, see also
\citealp{Burtscher2016}).

\subsection{Properties of clouds from eclipse events}

Cloud eclipse events provide constraints on the \textbf{angular size}
of individual clouds. In general, the clouds are observed to be very
small in angular extent subtended at the black hole ($\sim0.1'-1\text{°}$,
Nikutta et al., in prep.). However, observational biases due to limited
observing periods preferentially select clouds with high angular velocities
and small angular extents. It is furthermore reasonable to assume
that the sizes of clouds show a powerlaw-like distribution which spans
several orders of magnitude. In other words, we expect the observed
distribution to only cover the low-end tail of the intrinsic powerlaw
angular size distribution. For simplicity, we adopt an exponential
distribution of the angular sizes centred around $\theta^{{\rm cloud}}=1\text{°}$,
shown in the inset of Figure~\ref{fig:eclipses}. This encompasses
observed sizes (black histogram). The clouds are assumed to be spherical
blobs with diameter $D^{{\rm cloud}}=d\cdot\sin\theta^{{\rm cloud}}$
for a distance $d$.

Our model is also constructed to reproduce \textbf{eclipse event rates}.
The systematic study of cloud events by \citet{Markowitz2014} found
that the probability to see a cloud event increases with the number
of days between observations. Figure~\ref{fig:eclipses} indicates
detected events (error bars) and the sensitivity of their survey based
on their windowing function (steps), taken from their Figure~7. The
frequency of events in type 1 (cyan) and type 2 (orange) appear to
approximately follow a power law relation. To predict such observables
from our clumpy geometry, we assume for simplicity circular Keplerian
orbits on random planes of orientation. We set the orbital period
of the innermost cloud to one day, which for the outer-most clouds
translates to a period of eight years. The corresponding distances\footnote{The inner most cloud has a orbital radius of $42\,\mathrm{au}\,\left(M_{\mathrm{BH}}/10^{7}M_{\odot}\right)^{1/3}$.}
are consistent with observationally inferred distances \citep[approximately light days, see][for a review of constraints]{Markowitz2014}.
We then advance all orbits by a time $\Delta t$ (x-axis) and determine
what fraction of 20,000 random viewing sight-lines change column density
significantly (for type-1: unobscured become obscured, for type-2:
obscured at least double the column; obscured refers to the $\NH=10^{22}\mathrm{cm}^{-2}$
threshold). Because of the different sample selection, we do not assume
the same $\sigma$ values as in Figure~\ref{fig:Column-density-profile}.
Instead, we assume that type-2 detections are more likely to be created
in a more covered torus ($\sigma=40\text{°}$ gives the best fit)
than type-1 AGN ($\sigma=15\text{°}$). The slope and eclipse event
rates of type-1/type-2 AGN are reproduced, as Figure~\ref{fig:eclipses}
demonstrates. When reproducing the data of \citet{Markowitz2014}
in Figure~\ref{fig:eclipses}, we were free to chose the clumpy model
parameter $\sigma$ and the time normalisation. These control the
normalisation in the y and x-axis, respectively. The key result is
however that a powerlaw over sufficient orders of magnitudes is predicted,
with the correct slope. This result stems from the combination of
our assumed exponential angular size distribution and our wide range
in distances for the clouds ($Y=100$) leading to a wide range of
orbital and eclipse durations.

A visualisation of our model is presented in Figure~\ref{fig:Crosssections}.
This shows a cross-section through the vertical mid-plane for the
two models with angular widths $\sigma=15\text{°}$ and $40\text{°}$.
Table~\ref{tab:Model-parameters} in the appendix summarises the
assumptions and parameters of our clumpy model.

\subsection{Final model geometry with optional inner ring\label{subsec:Adding-a-inner}}

\begin{figure}
\begin{centering}
\includegraphics[width=0.3\columnwidth]{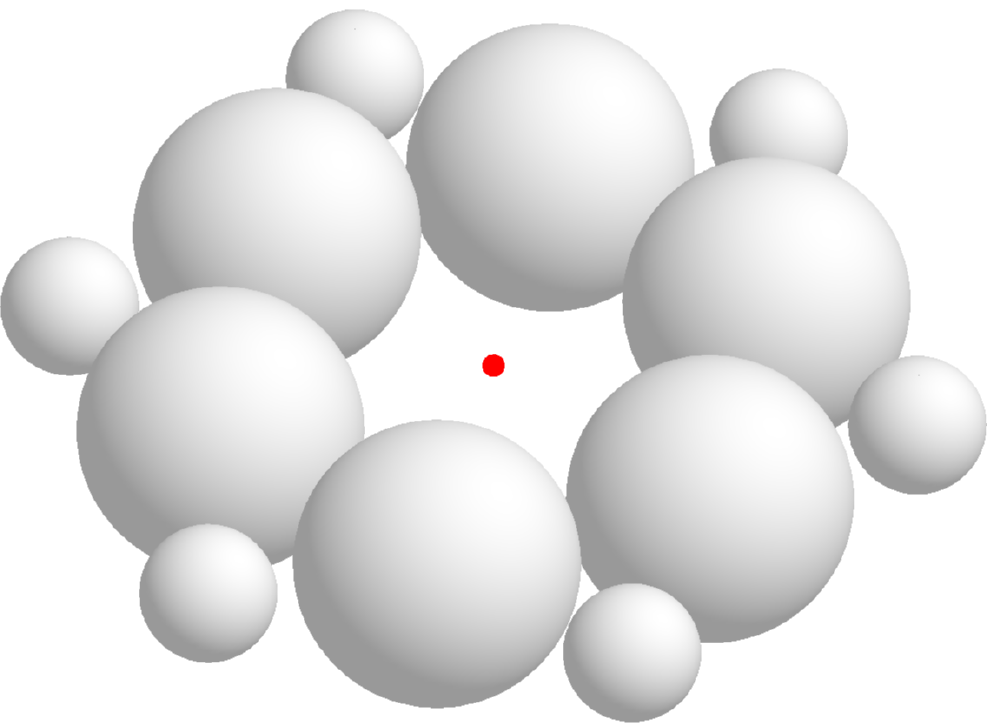}
\par\end{centering}
\caption{\label{fig:Compton-thick-cloud-ring}Compton-thick cloud ring around
central point source (red). Here six touching clouds are shown, with
a second row of gap-filling clouds. This configuration has a covering
factor of $30\%$. Our model allows variations in the covering factor
between $0\%$ (no cloud ring) and $60\%$ (3 touching inner clouds).}
\end{figure}

However, the clumpy model constructed above cannot fit some local
heavily obscured AGN. As discussed below and in more detail in a companion
paper, a highly covering, high-column density ($\NH\gtrsim10^{25}\mathrm{cm}^{-2}$)
reflector is necessary to reproduce the narrow high-energy Compton
hump of some local heavily obscured AGN, as traced by $8-10-20\mathrm{keV}$
X-ray colours (see Buchner et al., in prep). Briefly speaking, the
clumpy structure is composed of a wide range of column densities,
together producing a broad Compton hump. See Appendix~\ref{sec:profiles}
for further details regarding the spectrum of reflection off individual
clouds, as well as earlier work by \citet{Nandra1994a}. Introducing
an accretion disk that provides reflection to the powerlaw spectrum
proved ineffective, because both the intrinsic and reflected emission
is strongly attenuated by the torus. Instead, a reflecting mirror
is needed that has a virtually unobscured sight-line both to the corona
and the observer, while the sight-line between observer and corona
is obscured. To address this, we introduce into our clumpy model an
optional Compton-thick obscurer. While there are multiple equally
satisfactory geometries one could choose for this component, for computational
efficiency we compose it also of spheres. As illustrated in Figure~\ref{fig:Compton-thick-cloud-ring},
two rings of Compton-thick clouds, just touching each other, form
a thick donut. The covering fraction, $C$, determines the radius
and number of touching inner clouds used (three to sixteen). \emph{This
component remains optional so that its presence can be tested by parameter
fitting. }The discussion section §\ref{subsec:Interpretation-ring}
covers physical interpretations of this component.

We compute our final, two-component clumpy model grid across two parameters:
(1) the dispersion of the cloud population, \texttt{TORsigma} $\sigma$,
and (2) the covering factor of the Compton-thick inner ring, \texttt{CTKcover}
$C$, by changing the number of clouds. For computational efficiency,
and because the X-ray spectrum remains largely uneffected by this
choice, we reduce the number of clouds by a factor of ten to ${\cal N}_{{\rm tot}}=10^{5}$
and proportionally increase their angular area, setting $\theta^{{\rm cloud}}=\sqrt{10}\cdot1\text{°}\approx3\text{°}$.
Column density distributions and cloud eclipse predictions when this
component is included are shown in Figure~\ref{fig:Column-density-profile-1}
in the appendix. To achieve a smoother column density distribution
we also altered the cloud log-normal distribution to $N_{{\rm H}}^{{\rm cloud}}=10^{23.5\pm1.0}{\rm cm}^{-2}$.

\section{X-ray spectral model\label{sec:X-ray-spectral-model}}

\begin{figure}
\begin{centering}
\includegraphics[width=1\columnwidth]{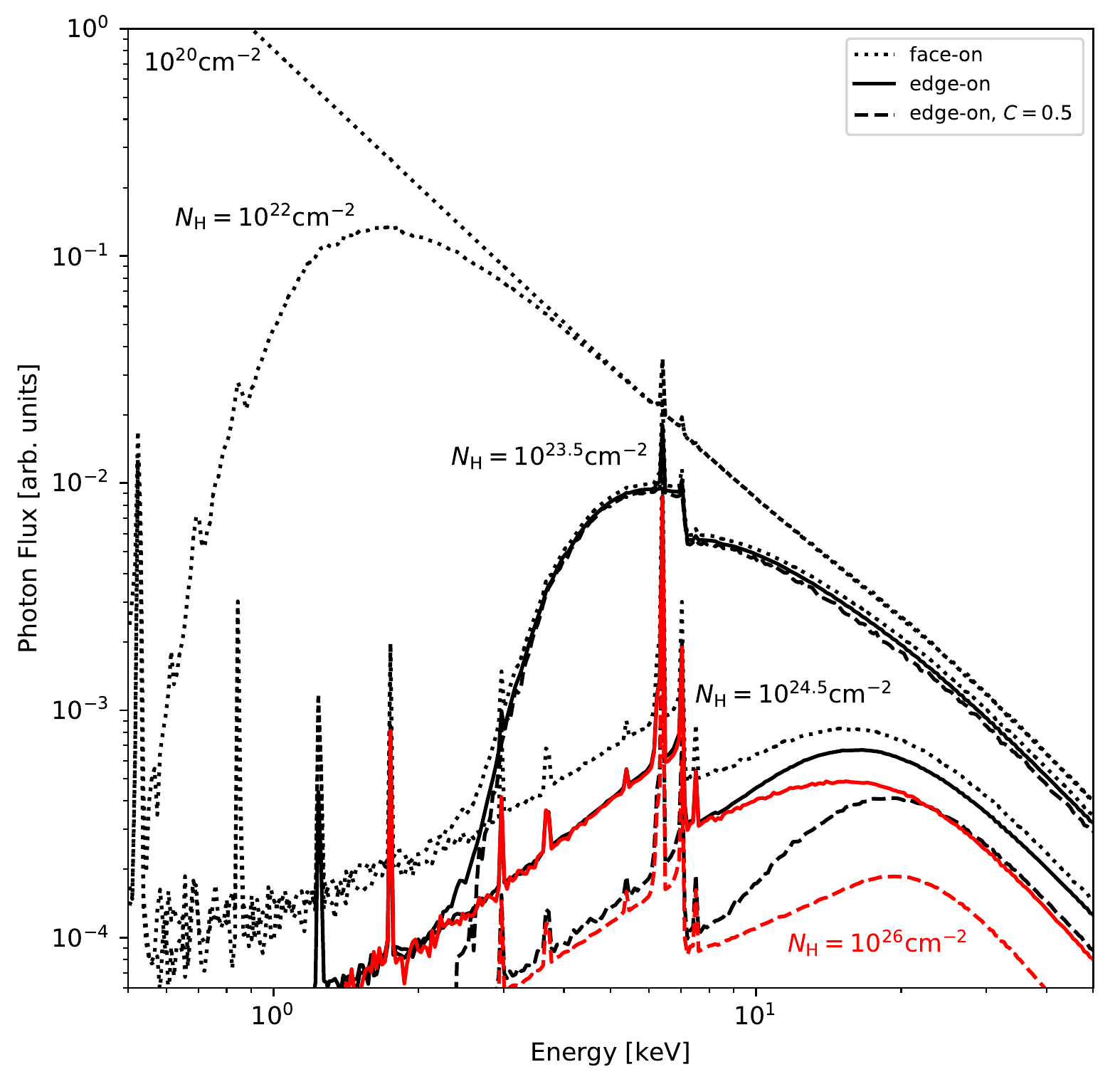}
\par\end{centering}
\caption{\label{fig:modelspectra}Clumpy model spectra. Viewing angles (dashed
\& solid lines) and LOS column densities (labelled at their respective
turn-overs energies) gives diversity in spectral shapes. The highest
column density, $\NH=10^{26}\mathrm{cm}^{-2}$ is shown in red. The
lines present typical face-on viewing angles (dotted lines, preferentially
low column densities, $0\text{°}$) and edge-on viewing angles (solid
and dashed lines, preferentially high column densities, $90\text{°}$).
The dashed lines have the inner ring described in §\ref{subsec:Adding-a-inner}
enabled.}
\end{figure}

\subsection{Model spectrum}

Figure~\ref{fig:modelspectra} presents various spectra produced
by one model geometry $\sigma=30\text{°}$. We input a powerlaw source
of photon index $\Gamma=2$ without a high-energy cut-off. Five LOS
column densities are shown, as well as two viewing angles. Towards
the Compton-thick regime, a scattering component appears and allows
soft photons to escape. Higher column densities also create a deeper
trough at $\sim10\mathrm{keV}$ relative to the Compton hump. The
trough also becomes deeper with $C$, as comparing solid and dashed
black curves reveals. In a companion paper, Buchner et al. (in prep.),
we explore this trough in detail with a $8-10-20\mathrm{keV}$ X-ray
colour-colour diagram. There we show that local Compton-thick AGN
show remarkably flat spectra at $8-10\mathrm{keV}$, similar to the
dashed red curve in Figure~\ref{fig:modelspectra}, which are difficult
to reproduce by existing models. We demonstrate this below in §\ref{sec:Obs}.

\subsection{X-ray spectral model release}

We release our spectral model, \uxclumpy{}\textsc{ (}Unified X-ray
\textsc{Clumpy} model), as an \texttt{\xspec{}} table model for download
at \href{https://doi.org/10.5281/zenodo.602282}{https://doi.org/10.5281/zenodo.602282}.
The parameters of the model are listed in Table~\ref{tab:Spectral-model-parameters}. 

The model can be used as follows in \texttt{\xspec{}} syntax: 

\begin{lstlisting}
model atable{uxclumpy.fits}
\end{lstlisting}
The model also allows for an exponential high-energy cut-off to more
realistically represent the X-ray continuum. This, and the addition
of other components, can make model fitting more degenerate and create
multiple local optima in parameter space. Therefore, we recommend
to initially freeze \texttt{Ecut} to the maximum value. Once a reasonable
fit has been obtained, constraining \texttt{Ecut} can be attempted.
Unrealistic fits show low photon indices ($\Gamma<1.4$) and low energy
cut-offs to mimic a Compton hump.

Compared to previous spectral models (e.g. \BNTORUS{}, \MYTORUS{}),
there are some important differences:

(1) \emph{The meaning of $\NH$ and viewing angle}: In our model,
the \texttt{NH} fitting parameter is the LOS column density. A given
geometry (\texttt{TORsigma}, \texttt{CTKcover}) produces a diversity
of LOS column densities. To understand the implications of the \texttt{NH}
and \texttt{Theta\_inc} parameters better, a few technicalities should
be mentioned: For each simulated photon escaping to infinity, its
direction is noted. For the same direction from the X-ray point source,
the LOS column density is computed. The photon is then placed in a
bin corresponding to column density and viewing angle (0-30°: face-on,
30-60°: intermediate, 60°-90°: edge-on). This binning scheme \textbf{partitions
the sky around the corona by column density and viewing angle}. In
practice, this makes spectral fitting easier, as the LOS column density
is the primary influence on spectral shape. Furthermore, physically
meaningful fits of column density variations over multiple observation
epochs are possible: Unlike existing models, it is not required that
the geometry (or equatorial torus column density) changes from source
to source, or from observation to observation, or that the viewing
angle changes drastically. Instead, multiple LOS column densities
are possible at very similar viewing angles, allowing realistic eclipse
events by slight rotation self-consistently in our model. We recommend
that multi-epoch fits keep the geometry constant, but allow variations
in \texttt{NH}.
\begin{table}
\caption{\label{tab:Spectral-model-parameters}Clumpy model spectral parameters.}

\centering{}%
\begin{tabular}{>{\centering}p{0.15\columnwidth}>{\centering}p{0.15\columnwidth}c>{\raggedright}p{0.33\columnwidth}}
Name & Symbol & Range & Description\tabularnewline
\hline 
\hline 
\texttt{PhoIndex} & $\Gamma$ & $1-3$ & Photon index\tabularnewline
\texttt{Ecut} & $E_{\text{cut}}$ & $60-400$  & Energy cut-off $[{\rm keV}]$\tabularnewline
\texttt{NH} & $\NH(\text{LOS})$ & $10^{20-26}$ & Total LOS column density $[{\rm cm}^{-2}]$\tabularnewline
\texttt{TORsigma} & $\sigma$ & $6-90\text{°}$ & Cloud dispersion\tabularnewline
\texttt{CTKcover} & $C$ & $0-0.6$ & Covering fraction of inner ring\tabularnewline
\texttt{Theta\_inc} & $\theta_{\text{inc}}$ & $0-90\text{°}$ & Viewing angle\tabularnewline
\end{tabular}
\end{table}

(2) \emph{Opening angle}s: A key geometric parameter of torus models
is the opening angle, $\theta_{{\rm op}}$. In our model, the inner
Compton-thick covering, \texttt{CTKcover}, can be directly interpreted
as a covering factor for that component. The cloud population has
however a fuzzy opening angle boundary, which is defined by the angular
width $\sigma$ from the plane where the LOS number of clouds decrease
to their $1\sigma$ value. The corresponding opening angle depends
on the considered column density and can be read off Figure~\ref{fig:Column-density-profile}.
Our angular width definition is consistent with the existing \textsc{Clumpy}
model definition (see~§\ref{sec:Infrared-Spectrum}). In the \BNTORUS{}
model, the opening angle $\Theta_{\text{tor}}$ can change slightly
the Compton hump of that model, and this has been used in some studies
to constrain the torus opening angle. In our model, the \texttt{CTKcover}
parameter provides a more powerful, geometrically consistent way to
probe the reflector geometry. Unfortunately, there exists no simple
geometry mapping between the models, as the best-matching photon index
is different at the corresponding LOS column densities.

\subsection{Warm mirror component\label{subsec:Warm-mirror-component}}

Absorbed AGN spectra frequently show excess soft photons. This can
be interpreted as a warm mirror: hot electrons Compton scattering
the intrinsic powerlaw past the obscurer \citep[e.g.][]{Matt2000a,Bianchi2006,Bianchi2010,Brightman2014}.
Frequently this component is simply modelled as a powerlaw with the
same photon index as the intrinsic powerlaw but a fraction of the
normalisation, following the Thomson scattering approximation. In
\texttt{\xspec{}}, this can be added with:
\begin{lstlisting}
model atable{uxclumpy.fits}+const*zpower
\end{lstlisting}
The parameters of the additional powerlaw should be set to the same
values as that of the \uxclumpy{} obscurer model.

However, if the warm mirror is volume-filling inter-clump hot gas
or the narrow-line region, the spectrum is probably more complex,
combining warm (hot gas) and cold reflection (dense clouds). As an
improved approximation, we sum the emission over all angles ($4\pi$)
for each geometry. To first order, this angle-averaged spectrum is
a powerlaw, because it is dominated by photons escaping without absorption.
To second order, a mild Compton hump and Fe K feature are present,
as in the unobscured sight-lines in Figure \ref{fig:modelspectra}.
To represent the warm mirror more accurately, we provide a second
table model that can be added instead of a simple powerlaw:

\begin{lstlisting}[breaklines=true]
model   atable{uxclumpy.fits} + 
  const*atable{uxclumpy-scattered.fits}
\end{lstlisting}
All table parameters of the angle-averaged table model (\texttt{uxclumpy-scattered},
warm reflected component) should be linked to their counterparts in
the obscurer model (\texttt{uxclumpy}, which includes transmitted
and cold reflected component). The relative normalisation constant
should not exceed $\sim10\%$ to avoid ambiguous powerlaw shapes and
unrealistic scattering efficiencies. Additional galactic and warm
absorption layers can be added on top as usual.

\begin{figure}
\includegraphics[width=1\columnwidth]{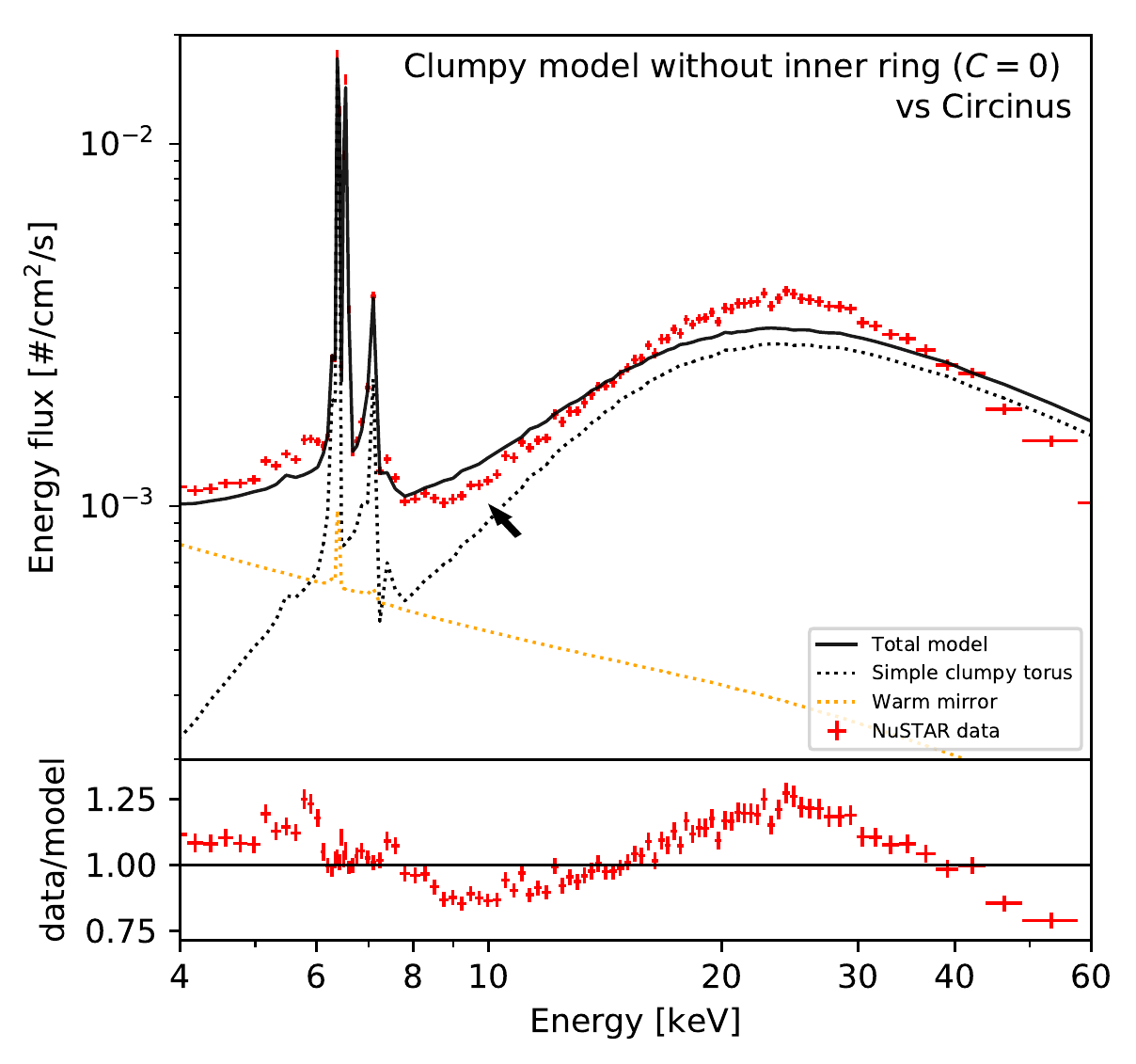}

\caption{\label{fig:misfit-Circinus}\emph{NuSTAR} observations of \emph{Circinus}
fitted with clumpy torus model without inner ring component. For visualisation,
the data are binned here to 1000 counts and deconvolved with the model.
Additionally, a warm mirror component (orange dotted line, see §\ref{subsec:Warm-mirror-component})
and lines at rest-frame energy $6.5$ and $7.1{\rm keV}$ were added
to the fit. The best fit has a LOS column density of $N_{{\rm H}}=3\times10^{24}{\rm cm}^{-2}$
and a photon index of $\Gamma=1.69$. This model rises already at
$\sim8{\rm keV}$ and produces a shallow Compton hump. In contrast,
the data remain flat until $10{\rm keV}$ before rising steeply. This
leads to substantial systematic residuals (positive at $6\mathrm{keV}$,
negative at $10\mathrm{keV}$, positive at $20-30\mathrm{keV}$) in
the \emph{lower panel}.}
\end{figure}

\begin{figure*}
\includegraphics[width=1\columnwidth]{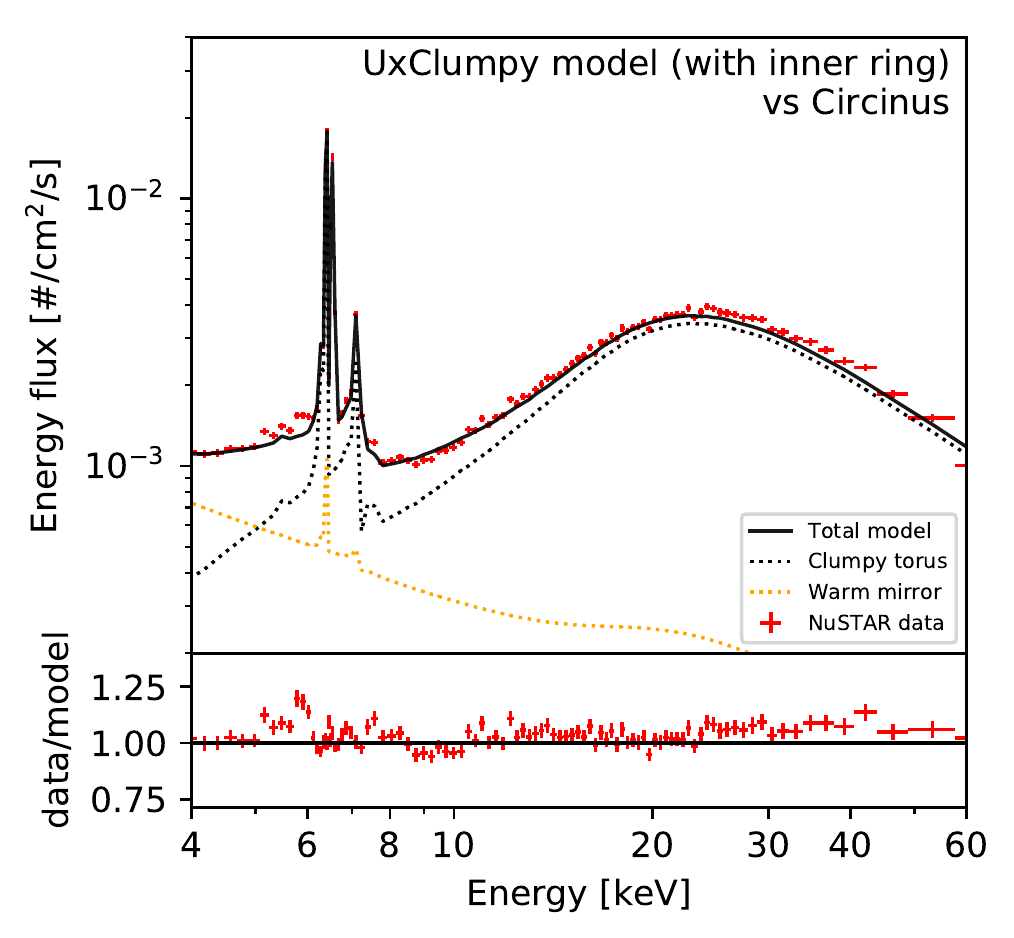}\includegraphics[width=0.95\columnwidth]{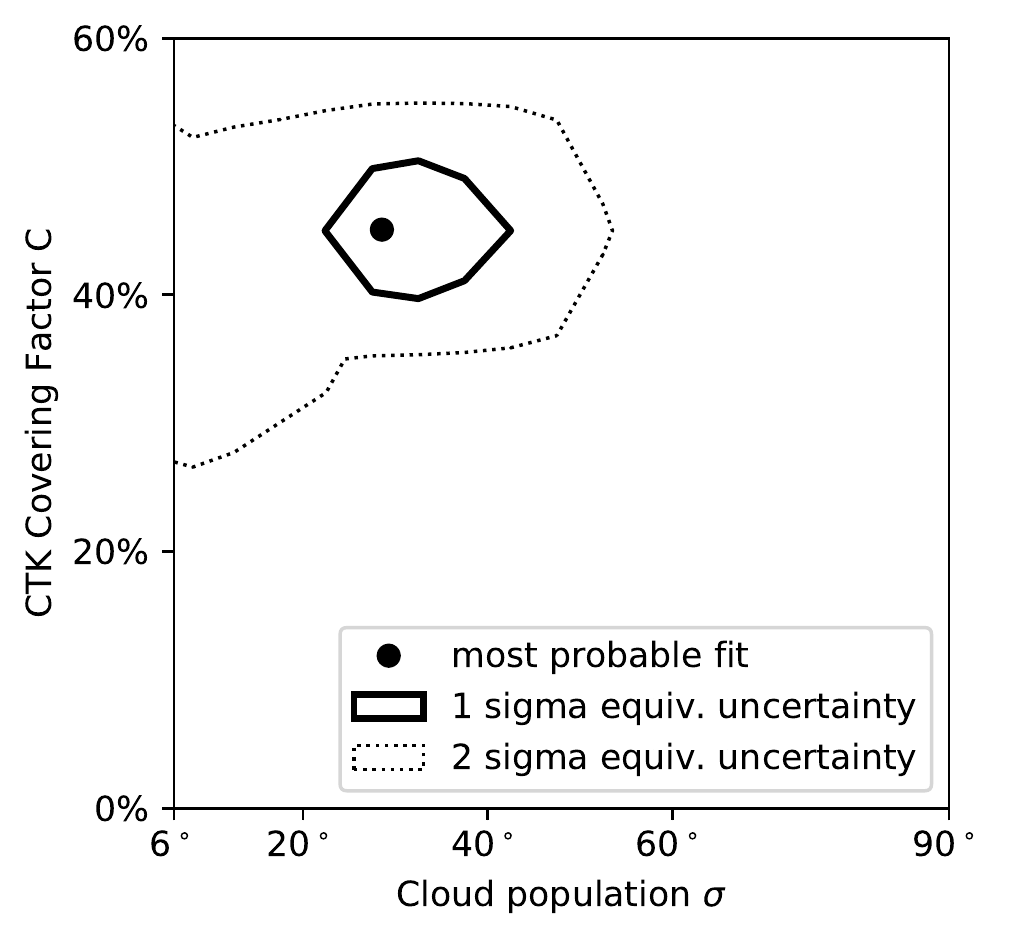}

\includegraphics[width=1\columnwidth]{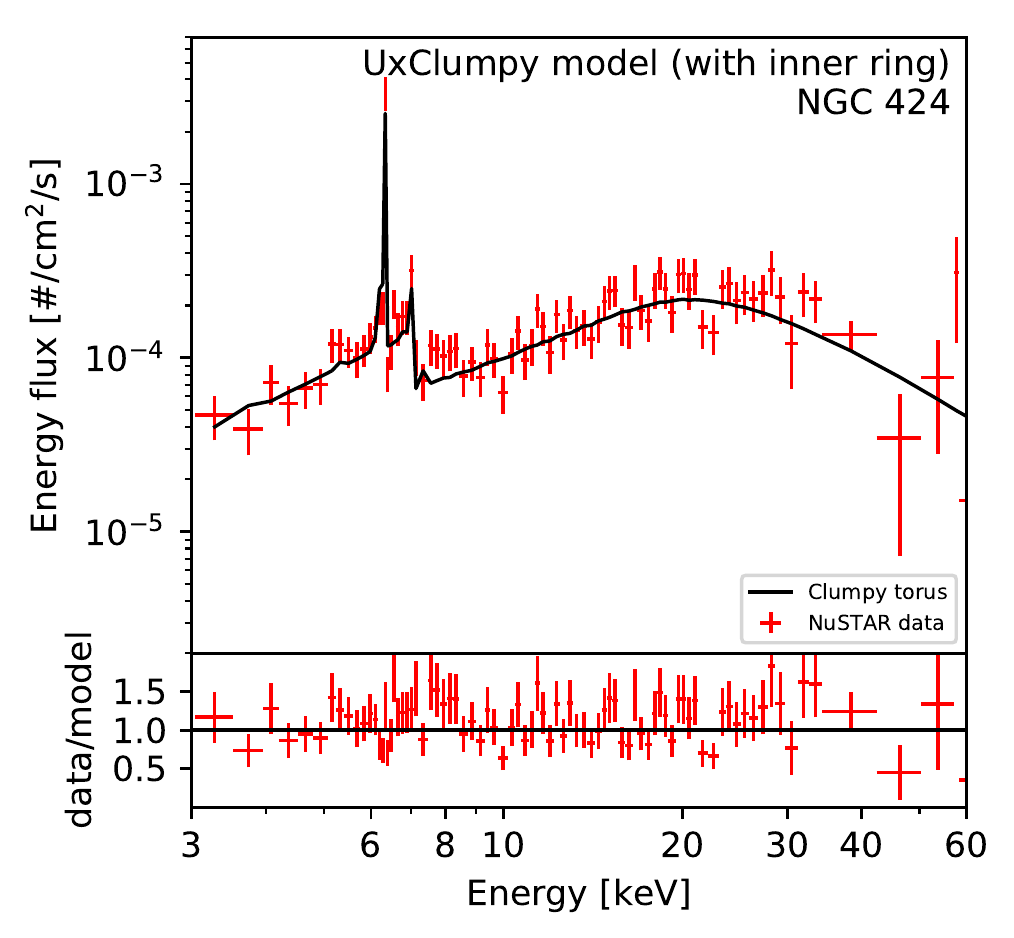}\includegraphics[width=0.95\columnwidth]{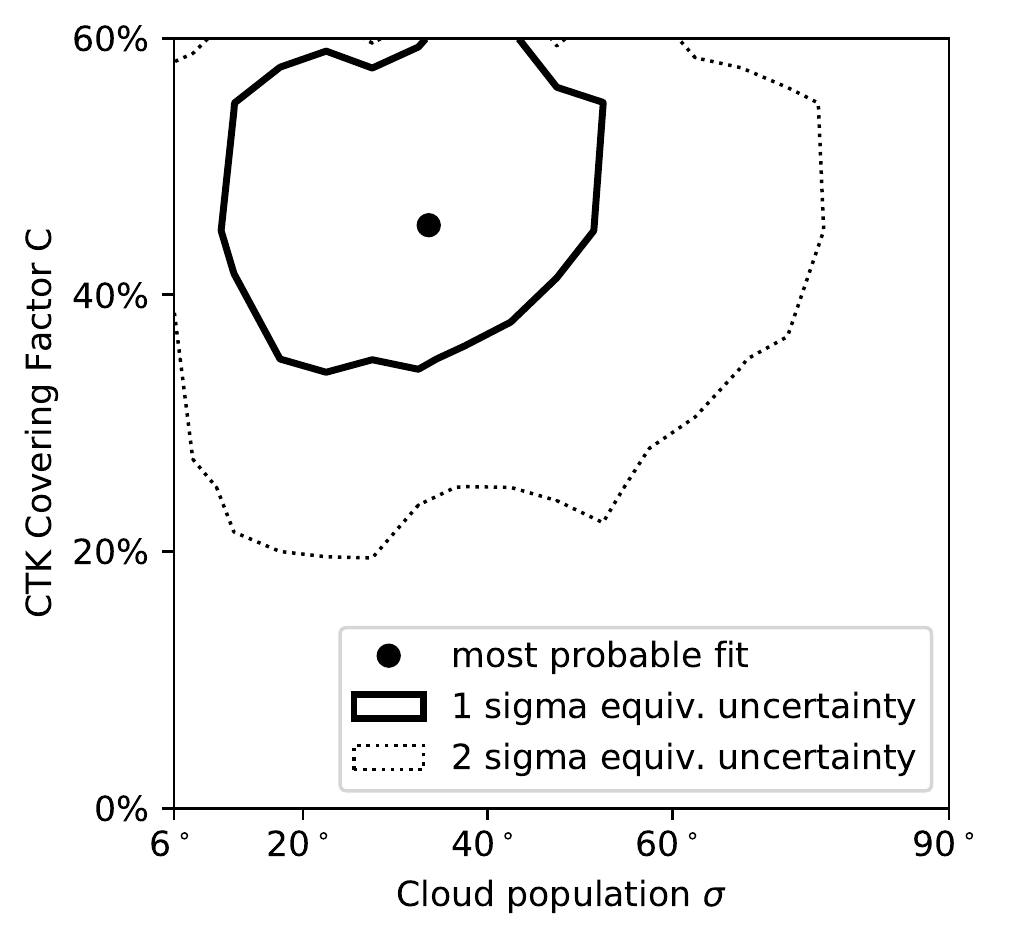}

\includegraphics[width=1\columnwidth]{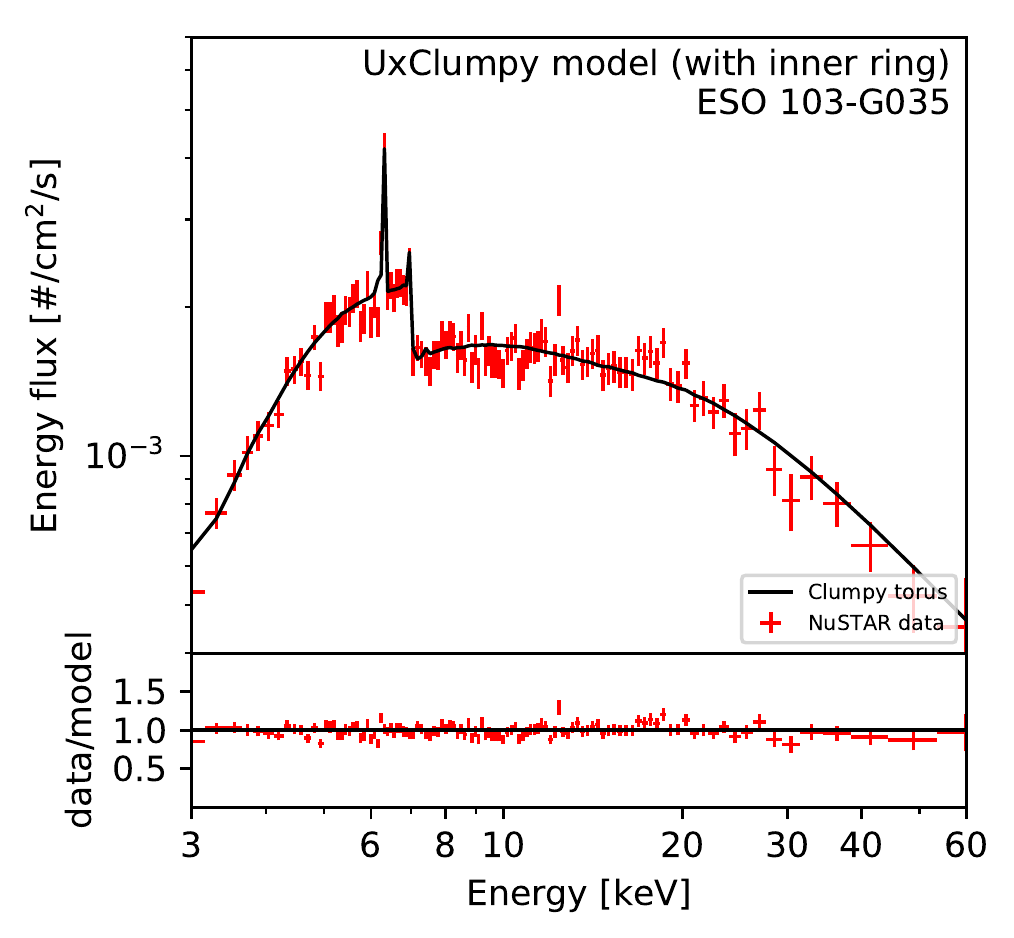}\includegraphics[width=0.95\columnwidth]{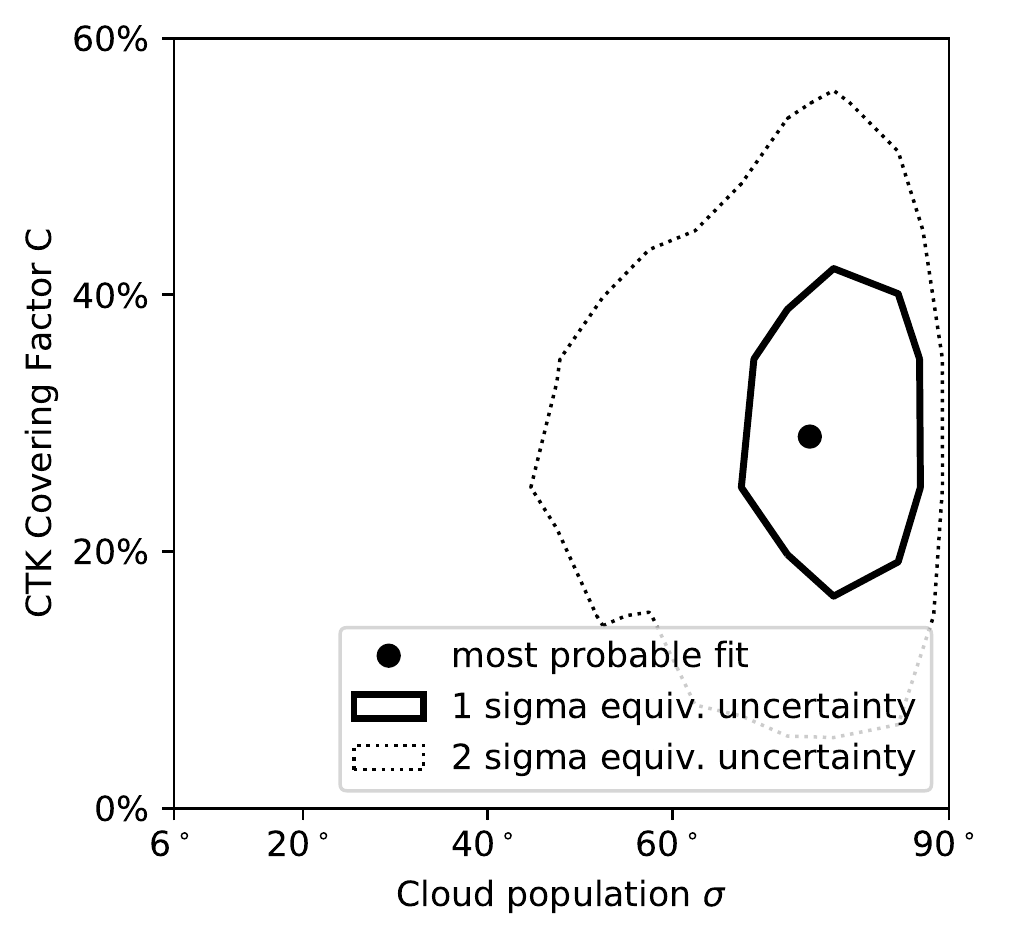}

\caption{\label{fig:Circinus2}\label{fig:NGC424}\label{fig:ESO103G035}\emph{Left
panels}: Clumpy torus model fit of Circinus, NGC~424 and ESO~103-G035
\emph{NuSTAR} observations, respectively. In each case a reasonable
fit (black line) is obtained with our unified clumpy torus model including
an inner ring (see §\ref{subsec:Adding-a-inner}). We added a Thomson
scattered component (see Section §\ref{sec:X-ray-spectral-model}
for details), which contributes substantially in Circinus. \emph{Right
panels} show constraints on the geometry parameters of the inner ring
(CTK covering factor, y-axis) and the cloud population ($\sigma$,
x-axis), within the allowed ranges of our model. High Compton-thick
covering factors are preferred for all three sources.}
\end{figure*}

\section{Fitting local Compton-thick AGN\label{sec:Obs}\label{subsec:Circinus:-Hard-X-ray}}

The nearby Seyfert 2 Circinus Galaxy (hereafter Circinus) is a prototypical
Compton-thick AGN, and due to its proximity has X-ray observations
of the nucleus with high signal-to-noise ratios and spatial resolution
\citep[see][for a recent analysis]{Arevalo2014}. The \emph{NuSTAR}
spectrum was extracted following standard procedures with the \emph{NuSTAR}
pipeline \citep[see][for details]{Brightman2015}.

First, we disable the inner ring component ($C=0$, see §\ref{subsec:Adding-a-inner}).
Figure~\ref{fig:misfit-Circinus} presents the best fit\footnote{We use the X-ray fitting tools \texttt{\xspec{}} \citep{XSPEC},
and \texttt{\sherpa{}} \citep{Freeman2001} with \texttt{\BXA{}}
\citep{Buchner2014}, bin spectra to 20 counts, and employ $\chi^{2}$
statistics.} of this simple clumpy model. In the \emph{NuSTAR} data the Compton
hump begins to rise at relatively high energies ($\sim10\,{\rm keV}$),
with the model producing a shallower Compton hump. This is caused
by the diversity of clouds creating Compton reflection: In low column
density clouds, singly and doubly Compton scattered emission can escape.
High column density clouds lead to self-absorption and a trough at
$8-15\,{\rm keV}$. In Appendix~\ref{sec:profiles} we discuss the
reflection spectra arising from single clouds in more detail. We find
similar results with other Compton-thick AGN (NGC~1068, NGC~424,
NGC~3393), and discuss this problem in more detail in a companion
paper, using a X-ray colour-colour diagnostic to characterise Compton
hump shapes. The spectral shape mismatch results in a poor fit ($\chi^{2}/\mathrm{dof}=2774/1703\approx1.6$).
A ``pure'' high-column density reflector dominating over all contamination
by lower column density reflectors (the majority of our current cloud
population) is needed. We anticipated the need for such a component
in §\ref{subsec:Adding-a-inner}. To reduce strong line residuals
distracting from the main point of Figure~\ref{fig:misfit-Circinus},
we included additional gaussian lines. While our model already includes
line emission, extended line emission from beyond the torus ($\gg$10pc)
is known in this source \citep{Arevalo2014}. Our results and conclusions
are consistent with and without these additional lines.

We now refit the data, with the inner ring component enabled. The
fit is much better ($\chi^{2}/\mathrm{dof}=1765/1714\approx1.0$,
top left panel in Figure~\ref{fig:Circinus2}), with the residuals
(lower panel) not showing systematic deviations. The best-fit has
a photon index of $\Gamma=2.11$, assuming no energy cut-off, behind
a line-of-sight column density of $\NH(\text{LOS})=4\times10^{24}{\rm cm}^{-2}$.
Because the \textit{NuSTAR} extraction region exhibits substantial
contamination by other X-ray sources \citep{Arevalo2014}, we now
focus on fitting only the $\geq8\,{\rm keV}$ energy range to derive
physical properties of the reflector. In this case, we find a photon
index of $\Gamma=\paramCircinusphoindex$, behind a line-of-sight
column density of $\NH(\text{LOS})=\paramCircinusnhHiLo{\rm cm}^{-2}$.
The high-energy cutoff uncertainties, $E_{cut}=\paramCircinusecut$,
are degenerate with those of $\Gamma$. Importantly, we are able to
constrain the geometry of the obscurer (contours in top right panel
of Figure~\ref{fig:Circinus2}). The clumpy component has a covering
angle $\sigma$ below $45\text{°}$, and the Compton-thick covering
factor of the inner ring is $C>40\%$, that is, the corona is extensively
covered by Compton-thick material.

To study the soft energy AGN emission as well, we consider NGC~424.
Other nearby Compton-thick AGN (such as Circinus) are often contaminated
by nearby sources in the host galaxy, such as supernova remnants,
ultraluminous X-ray sources and X-ray binaries. NGC~424 is much more
luminous than Circinus and therefore its nuclear emission dominates
over such host galaxy contaminants. The soft energies show no contamination
in \emph{XMM-Newton} observations down to energies of $\sim3{\rm keV}$
\citep{Marinucci2011}. The middle left panel of Figure~\ref{fig:NGC424}
presents our best fit to \emph{NuSTAR} spectra in the $3-65{\rm keV}$
range. The fit is of good quality ($\chi^{2}/\mathrm{dof}=84/79\approx1.1$)
and gives a photon index of $\Gamma=\paramNGCfourphoindex$ without
the need for a high-energy cutoff (parameter unconstrained). Such
a high photon index is expected from this high-Eddington accretion
rate source ($L_{\text{bol}}/L_{\text{Edd}}\approx0.1$, \citealp{Bian2007,Brightman2013,Brightman2017}).
This X-ray emitter is behind a line-of-sight column density of $\NH(\text{LOS})=\paramNGCfournhHiLo{\rm cm}^{-2}$.
We are able to constrain the geometry of the obscurer (contours in
the centre right panel of Figure~\ref{fig:Circinus2}): The clumpy
component has a covering angle $\sigma<60\text{°}$, and the Compton-thick
covering factor of the inner ring is $C>30\%$.

We further consider ESO~103-G035, a Compton-thin source. The \emph{NuSTAR}
spectrum of this source is soft without strong obscuration or host
contamination, therefore we consider the entire \emph{NuSTAR} spectrum
($3-70\,{\rm keV}$). The bottom left panel of Figure~\ref{fig:ESO103G035}
shows our best fit for this source, which is again of good quality
($\chi^{2}/\mathrm{dof}=978.58/1020\approx0.96$) and gives a photon
index of $\Gamma=\paramESOphoindex$ with a high-energy cutoff $E_{cut}=\paramCircinusecut$.
This X-ray emitter is behind a line-of-sight column density of $\NH(\text{LOS})=\paramESOnh{\rm cm}^{-2}$.
We are also able to constrain the geometry of the obscurer in this
case (contours in the bottom right panel of Figure~\ref{fig:ESO103G035}):
The clumpy component has a covering angle $\sigma>60\text{°}$ under
a face-on viewing angle and the Compton-thick covering factor of the
inner ring is $C=\paramESOctkcover\%$.

All three spectra prefer large covering fractions of the high-column
density inner ring reflector. In Compton-thick AGN the cloud population,
by means of low $\sigma$ values, tends to be hidden away in the shadow
of this inner ring, avoiding low-column density reflection. Indeed,
the emerging spectrum and fits change insignificantly when the torus
clumps are removed ($\sigma=0$). In Compton-thin AGN, the cloud population
provides the Compton-thin line-of-sight absorber, while the inner
ring is again significantly needed as a Compton-thick reflector.

\section{Discussion}

We have presented a clumpy torus model geometry, \uxclumpy{}\textsc{,}
from which we simulate X-ray spectra. \uxclumpy{} provides (1) a
wide coverage of energies ($0.1-1000\mathrm{keV}$), (2) a wide range
of LOS column densities ($10^{20-26}\mathrm{cm}^{-2}$), (3) a variable
energy cut-off and (4) fluorescent emission lines self-consistently.
Additionally, the clumps in our geometry reproduce the frequency,
duration and sizes of eclipse events and X-ray eclipse event observations
can be modelled self-consistently without requiring a drastic change
in the viewing angle.

The spectral shapes around the Compton hump and trough above the Fe
edge differ for the various geometries and viewing angles, and are
diverse (see Figure~\ref{fig:modelspectra}). We additionally compute
a better approximation of the warm mirror component. Overall however
\uxclumpy{} produces spectral shapes similar to the \BNTORUS{} model,
which has been found to fit observations well. The added benefit of\textsc{
}\uxclumpy{} is the ability to infer novel physical characteristics
from the spectral fits and consider the energy cut-off. Below we discuss
the multi-wavelength spectrum of our model and interpret our inner
ring component.

\subsection{Multi-wavelength Spectrum\label{sec:Infrared-Spectrum}}

\begin{figure*}
\includegraphics[width=1\textwidth]{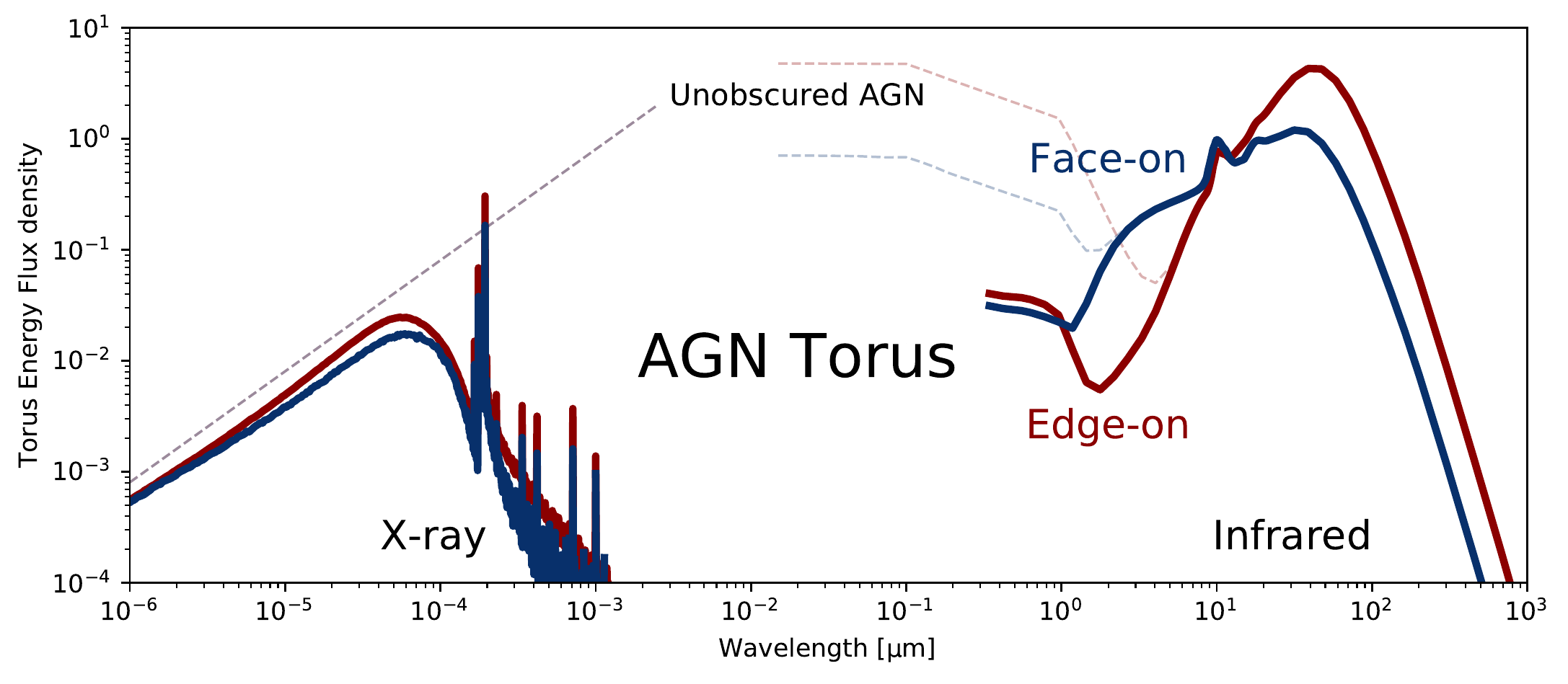}\caption{\label{fig:irspectrum}Multiwavelength spectrum of our unified clumpy
geometry. The clumpy torus with $\sigma=30\text{°}$ (left panel of
\ref{fig:Crosssections}) is seen through a obscured, Compton-thick
line-of-sight either edge-on (red) or face-on (blue). Dashed lines
indicate the case of a unobscured view of the accretion disk. }
\end{figure*}

\begin{table*}
\caption{\label{tab:Model-parameters}Model parameters.}

\begin{tabular}{ccccc}
Description & Distribution & Parameter & Value & Determined from\tabularnewline
\hline 
Total number of clouds & fixed & ${\cal N}_{{\rm tot}}$ & $10^{5}$ & Column density distribution\tabularnewline
Vertical distribution & Gaussian & $m$ & 2 & Column density distribution\tabularnewline
Vertical dispersion & variable & $\sigma$ & 6°- 90° & Column density distribution\tabularnewline
Radial distribution  & uniform & $q$ & 0 & Assumed\tabularnewline
Radial extent & fixed & $Y$ & 100 (X-ray), 10 (IR) & Assumed\tabularnewline
\hline 
Cloud sizes & Exponential & $\theta^{{\rm cloud}}$ & $1\text{°}$ & Eclipsing events\tabularnewline
Column density of clouds & Log-Normal & $N_{{\rm H}}^{{\rm cloud}}$ & $10^{23.5\pm1.0}$ & Column density distribution\tabularnewline
\hline 
Number of clouds (plane) & (from above) & ${\cal N}_{0}$ & $2-9$ & Computed\tabularnewline
Volume filling factor & (from above) & $f_{V}$ & $2-5\%$ & Computed\tabularnewline
Cloud mean optical depth & (from above) & $\tau_{V}$ & $150$ & Assuming Galactic relation\tabularnewline
\end{tabular}
\end{table*}
Infrared and X-ray data can be self-consistently analysed under our
clumpy geometry model, by construction. The emerging infrared model
was computed using \model{CLUMPY} \citep{Nenkova2008a,Nenkova2008}
and is depicted in Figure~\ref{fig:irspectrum}. \model{CLUMPY}
performs probabilistic ultraviolet-optical-infrared radiative transfer.
This approximation requires that the volume filling factor is small
($<10\%$) and the cloud sizes are small, which is the case given
our assumed small angular sizes.

\begin{figure*}
\centering{}\includegraphics[width=1\textwidth]{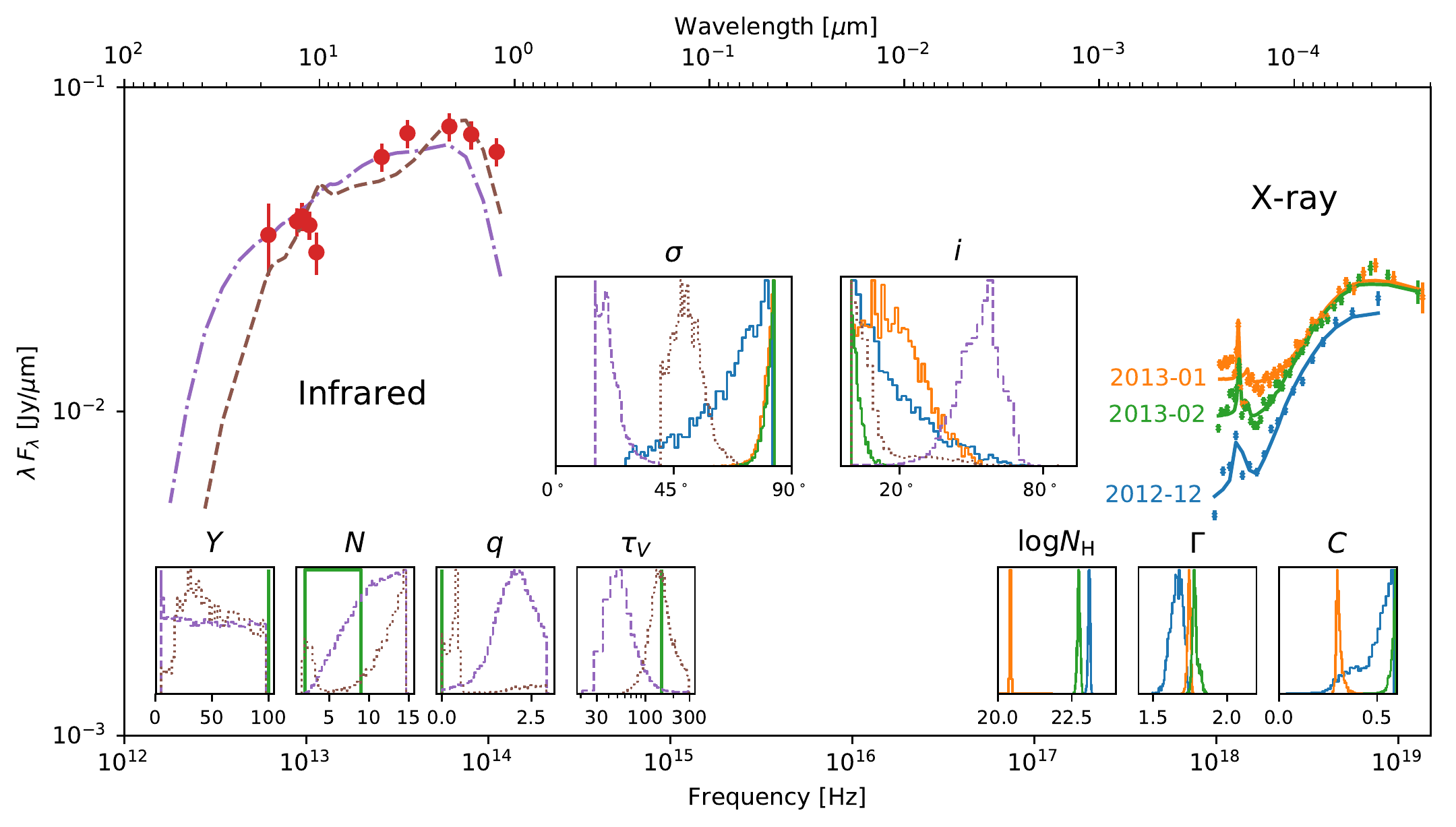}\caption{\label{fig:irfit}Multiwavelength fit of our unified clumpy geometry
to data of NGC~1365. Infrared photometry on the left overlayed with
two \model{CLUMPY} fit solutions, and spectra from three \emph{NuSTAR}
observations are shown on the right with \uxclumpy{} fits. Insets
show parameter probability distributions for each fitted dataset in
corresponding colours. The infrared fit solution shown in brown is
overall compatible with the X-ray fits (blue, green, orange).}
\end{figure*}

Table~\ref{tab:Model-parameters} presents the model parameters and
assumptions that went into constructing our clumpy cloud population.
For a self-consistent physical model in the infrared, the \textsc{Clumpy}
torus parameters should be set as follows (see Table~\ref{tab:Model-parameters}):
$Y\approx10-20$ ($<100$), $\sigma={\rm TORsigma}$, $N_{0}=2-5$
(depending mildly on $\sigma$), $q=0$, and $\tau_{{\rm V}}\approx120-160$. 

Figure~\ref{fig:irspectrum} presents the X-ray model spectrum with
the reprocessed infrared emission from the . We normalised at $12{\rm \mu m}$
and use the empirical ratio of 1.7 between the $12{\rm \mu m}$ and
intrinsic $2-10\,{\rm keV}$ luminosity \citep[e.g.][]{Asmus2015}.
In the optical and ultraviolet the intrinsic AGN disk emission dominates
along unobscured sight-lines (dashed lines in Figure~\ref{fig:irspectrum}).
At long wavelengths, the reprocessed emission is gray-body-like and
dominated by the coldest dust regions. In the mid-infrared, self-absorption
in edge-on systems hides the hot inner dust emission, while face-on
views expose a range of temperatures. The silicate feature is visible
around $10$ and $18{\rm \mu m}$. In the X-ray regime very energetic
photons can penetrate the obscurer and escape. Between $\sim0.2-1\mathrm{nm}$
a tail of soft photons circumvents the line-of-sight obscurer by being
reflected off distant material. In addition to the continuuum, K shell
fluorescence features are clearly visible, the most prominent being
Fe~K.

We now attempt a multi-wavelength fit of the obscurer from X-rays
to infrared. To illustrate the clumpy obscurer, we select NGC~1365,
notorious for changing its LOS column density several times over time-scales
of days to years \citep[e.g.,][]{Risaliti2005,Risaliti2007,Rivers2015}.
A successful model should allow the obscurer geometry and inclination
to stay the same with only small changes in the sight-line to the
corona. Simultaneously, we expect the same geometry to also fit the
infrared photometry. Archival photometric data \citep[we imposed a 10\% minimum flux uncertainty]{McAlary1983,Skrutskie2006,RamosAlmeida2009,Asmus2014}
and \emph{NuSTAR} spectra \citep[ObsIDs 60002046005,7,9; 5-77 keV]{Rivers2015}
are presented in Figure~\ref{fig:irfit}. In the three observations
(blue, orange, green) in the left of Figure~\ref{fig:irfit}, the
spectra differ noticibly in shape. In the insets, the \uxclumpy{}
parameter probability distributions of a fit to each spectrum are
shown with corresponding colours. Fitting the infrared photometry
with \model{CLUMPY} leads to two solutions, shown in brown dashed
and purple dot-dashed curves. The parameter distributions are again
shown in the insets. We compare the solution shown in brown, with
a face-on inclination $i<20{^\circ}$, a substantial torus thickness
$\sigma>30{^\circ}$ (see central insets of Figure~\ref{fig:irfit})
to the constraints from the X-ray fits (or assumed \uxclumpy{} parameters
shown in the right insets). Overall, the derived parameters are compatible.
This allows the interpretation that a clumpy structure surrounds NGC~1365
whose inner-most clouds vary the X-ray column density and whose outer,
dusty clouds irradiate in the infrared. A single geometry, with the
same inclination angle was able to simultaneously reproduce column
density variations over time and the broad-band X-ray to infrared
spectral energy distribution. However, the presented fit does not
model all of the complexities present in the data. For example, the
least obscured spectrum (orange) shows complex, warm absorption \citep[see][]{Rivers2015}.
Additionally, in other variable-obscuration sources we find that a
multi-wavelength fit gives incompatible geometry constraints, perhaps
indicating that the structures differ. This indicates that further
modelling is needed and will be addressed in future work.

\subsection{Interpretation of the inner ring component\label{subsec:Interpretation-ring}}

When the inner ring component is present, the X-ray spectrum around
the Compton hump is similar to pure cold reflection (e.g. off an infinitely
thick slab, or a smooth donut). In the literature such a bulged component
has been proposed and interpreted for a variety of reasons. It could
be a Compton-thick cloud wind launch site \citep{Krolik1988}, an
inner wall \citep{Lightman1988} or a puffed up inner rim of a hot
disk as seen in proto-planetary disks \citep{Dullemond2001,Dullemond2010}.
The spectral shapes are also similar to the spectrum of a warped disk
(see Buchner et al., in prep), and can also be interpreted as one.
This is probably the most suitable interpretation for Circinus, where
an edge-on maser disk is seen \citep{Greenhill2003}. In the companion
paper Buchner et al. (in prep), a warped disk model provides the best
fit.

For compatibility with \model{CLUMPY} IR models, we can assume that
the inner Compton-thick reflector is dust-free and does not contribute
significantly to the infrared spectrum \citep[see simulations of][for a component of similar column density]{Gallagher2015}.
In this case, if it is interpreted as part of a dust-free broad emission
line region, very smooth line profiles do not necessitate that it
has further discrete clumpy substructure \citep[e.g.,][]{Laor2006}.
Alternatively, a similar inner ring component was introduced by \citet{Pier1993}
to explain the observed, hot ($\sim1300{\rm K}$) dust component \citep[see][for a review]{Netzer2015};
see also \citet{Kishimoto2011} and \citet{Hoenig2012} who considered
transferring the ``puffed-up inner rim'' model from proto-planetary
disks \citep{Dullemond2001,Dullemond2010} to AGN. 

Recent interferometric observations in the mid-infrared indicate that
the dust emission is in several sources primarily polar with respect
to the disk \citep{Asmus2016,Tristram2014}. Incorporation of these
results goes beyond this work. However, clumpy geometries that address
these can be calculated rapidly with our \texttt{XARS} code.

\section{Conclusion}

We constructed a clumpy model for the obscurer of AGN which reproduces
1) the column density distribution of the AGN population and 2) observed
cloud eclipse events. However, when applying the clumpy model to several
nearby, heavily obscured AGN, we find differences between the observed
\emph{NuSTAR} spectra and the model, in particular in the energy region
between the Fe K edge and the Compton hump. To resolve this, it is
necessary to insert close to the corona a Compton-thick reflector
with a substantial covering factor. This new inner component could
be interpreted as an inner rim, or a warped or bulged disk. It cannot
be attributed to reflection off a simple disk because the reflector
must simultaneously block the line of sight to the central engine
and reflect it. 

The need for the inner Compton-thick reflector (from X-ray spectral
fits) and the need for the clumpy component (from eclipses and infrared
observations) together could imply three classes of AGN of increasing
inclination angle: 1) Unobscured; 2) Obscured by clumps; and 3) Heavily
obscured by the inner reflector. The former two can have strong Compton
reflection from the inner ring and show eclipses. The latter is reflection-dominated
and may produce Compton-thick column density variations \citep[as seen for instance in NGC 1068,][]{Marinucci2016}.

We have developed a powerful Monte Carlo X-ray radiative transfer
code, \texttt{XARS}, which can deal with arbitrary geometries. We
release this open source software at \url{https://github.com/JohannesBuchner/xars}.

We release our \uxclumpy{} model as \texttt{\xspec{}} tables available
at \href{https://doi.org/10.5281/zenodo.602282}{https://doi.org/10.5281/zenodo.602282}.
The spectral model is similar in its ease of use, parameters and spectral
shape as the popular \BNTORUS{} model (see §\ref{sec:X-ray-spectral-model}).
It also allows a physical interpretation of the underlying geometry
including covering factors and enables self-consistent analysis of
variable column densities \citep[see also][]{Simm2018}. We also note
that our geometry has corresponding \model{CLUMPY} infrared spectra\footnote{Available at \href{https://www.clumpy.org}{https://www.clumpy.org}},
paving the way for self-consistent multi-wavelength analyses.

\section{Acknowledgements}

We thank the anonymous referee for careful reading of the manuscript
and their insightful comments. JB thanks Marc Schartmann and Leonard
Burtscher for discussions on tori in the infrared. JB thanks Eva Lefa
for sharing her Monte Carlo code to verify our Compton scattering
implementation. JB thanks Mislav Balokovi\'{c} for insightful conversations.

We acknowledge support from the CONICYT-Chile grants Basal-CATA PFB-06/2007
\& AFB-170002 (JB, FEB), FONDECYT Regular 1141218 (FEB), FONDECYT
Postdoctorados 3160439 (JB) and the Ministry of Economy, Development,
and Tourism's Millennium Science Initiative through grant IC120009,
awarded to The Millennium Institute of Astrophysics, MAS (JB, FEB).
This research was supported by the DFG cluster of excellence ``Origin
and Structure of the Universe''.

\appendix

\bibliographystyle{aa}
\bibliography{agn}

\begin{thebibliography}{80}
\expandafter\ifx\csname natexlab\endcsname\relax\def\natexlab#1{#1}\fi

\bibitem[{{Aird} {et~al.}(2015){Aird}, {Coil}, {Georgakakis}, {Nandra},
  {Barro}, \& {Perez-Gonzalez}}]{Aird2015}
{Aird}, J., {Coil}, A.~L., {Georgakakis}, A., {et~al.} 2015, ArXiv e-prints

\bibitem[{{Anders} \& {Grevesse}(1989)}]{Anders1989}
{Anders}, E. \& {Grevesse}, N. 1989, \gca, 53, 197

\bibitem[{{Antonucci}(1993)}]{Antonucci1993}
{Antonucci}, R. 1993, \araa, 31, 473

\bibitem[{{Ar{\'e}valo} {et~al.}(2014){Ar{\'e}valo}, {Bauer}, {Puccetti},
  {Walton}, {Koss}, {Boggs}, {Brandt}, {Brightman}, {Christensen}, {Comastri},
  {Craig}, {Fuerst}, {Gandhi}, {Grefenstette}, {Hailey}, {Harrison}, {Luo},
  {Madejski}, {Madsen}, {Marinucci}, {Matt}, {Saez}, {Stern}, {Stuhlinger},
  {Treister}, {Urry}, \& {Zhang}}]{Arevalo2014}
{Ar{\'e}valo}, P., {Bauer}, F.~E., {Puccetti}, S., {et~al.} 2014, \apj, 791, 81

\bibitem[{{Arnaud}(1996)}]{XSPEC}
{Arnaud}, K.~A. 1996, in Astronomical Society of the Pacific Conference Series,
  Vol. 101, Astronomical Data Analysis Software and Systems V, ed. G.~H.
  {Jacoby} \& J.~{Barnes}, 17

\bibitem[{{Asmus} {et~al.}(2015){Asmus}, {Gandhi}, {H{\"o}nig}, {Smette}, \&
  {Duschl}}]{Asmus2015}
{Asmus}, D., {Gandhi}, P., {H{\"o}nig}, S.~F., {Smette}, A., \& {Duschl}, W.~J.
  2015, \mnras, 454, 766

\bibitem[{{Asmus} {et~al.}(2016){Asmus}, {H{\"o}nig}, \& {Gandhi}}]{Asmus2016}
{Asmus}, D., {H{\"o}nig}, S.~F., \& {Gandhi}, P. 2016, \apj, 822, 109

\bibitem[{{Asmus} {et~al.}(2014){Asmus}, {H{\"o}nig}, {Gandhi}, {Smette}, \&
  {Duschl}}]{Asmus2014}
{Asmus}, D., {H{\"o}nig}, S.~F., {Gandhi}, P., {Smette}, A., \& {Duschl}, W.~J.
  2014, \mnras, 439, 1648

\bibitem[{{Balokovi{\'c}} {et~al.}(2018){Balokovi{\'c}}, {Brightman},
  {Harrison}, {Comastri}, {Ricci}, {Buchner}, {Gandhi}, {Farrah}, \&
  {Stern}}]{Balokovic2018}
{Balokovi{\'c}}, M., {Brightman}, M., {Harrison}, F.~A., {et~al.} 2018, ArXiv
  e-prints

\bibitem[{{Bauer} {et~al.}(2015){Bauer}, {Ar{\'e}valo}, {Walton}, {Koss},
  {Puccetti}, {Gandhi}, {Stern}, {Alexander}, {Balokovi{\'c}}, {Boggs},
  {Brandt}, {Brightman}, {Christensen}, {Comastri}, {Craig}, {Del Moro},
  {Hailey}, {Harrison}, {Hickox}, {Luo}, {Markwardt}, {Marinucci}, {Matt},
  {Rigby}, {Rivers}, {Saez}, {Treister}, {Urry}, \& {Zhang}}]{Bauer2014}
{Bauer}, F.~E., {Ar{\'e}valo}, P., {Walton}, D.~J., {et~al.} 2015, \apj, 812,
  116

\bibitem[{{Bian} \& {Gu}(2007)}]{Bian2007}
{Bian}, W. \& {Gu}, Q. 2007, \apj, 657, 159

\bibitem[{{Bianchi} {et~al.}(2010){Bianchi}, {Chiaberge}, {Evans}, {Guainazzi},
  {Baldi}, {Matt}, \& {Piconcelli}}]{Bianchi2010}
{Bianchi}, S., {Chiaberge}, M., {Evans}, D.~A., {et~al.} 2010, \mnras, 405, 553

\bibitem[{{Bianchi} {et~al.}(2006){Bianchi}, {Guainazzi}, \&
  {Chiaberge}}]{Bianchi2006}
{Bianchi}, S., {Guainazzi}, M., \& {Chiaberge}, M. 2006, \aap, 448, 499

\bibitem[{{Brightman} {et~al.}(2017){Brightman}, {Balokovi{\'c}}, {Ballantyne},
  {Bauer}, {Boorman}, {Buchner}, {Brandt}, {Comastri}, {Del Moro}, {Farrah},
  {Gandhi}, {Harrison}, {Koss}, {Lanz}, {Masini}, {Ricci}, {Stern},
  {Vasudevan}, \& {Walton}}]{Brightman2017}
{Brightman}, M., {Balokovi{\'c}}, M., {Ballantyne}, D.~R., {et~al.} 2017, ArXiv
  e-prints

\bibitem[{{Brightman} {et~al.}(2015){Brightman}, {Balokovi{\'c}}, {Stern},
  {Ar{\'e}valo}, {Ballantyne}, {Bauer}, {Boggs}, {Craig}, {Christensen},
  {Comastri}, {Fuerst}, {Gandhi}, {Hailey}, {Harrison}, {Hickox}, {Koss},
  {LaMassa}, {Puccetti}, {Rivers}, {Vasudevan}, {Walton}, \&
  {Zhang}}]{Brightman2015}
{Brightman}, M., {Balokovi{\'c}}, M., {Stern}, D., {et~al.} 2015, \apj, 805, 41

\bibitem[{{Brightman} \& {Nandra}(2011{\natexlab{a}})}]{Brightman2011a}
{Brightman}, M. \& {Nandra}, K. 2011{\natexlab{a}}, \mnras, 413, 1206

\bibitem[{{Brightman} \& {Nandra}(2011{\natexlab{b}})}]{Brightman2011b}
{Brightman}, M. \& {Nandra}, K. 2011{\natexlab{b}}, \mnras, 414, 3084

\bibitem[{{Brightman} {et~al.}(2014){Brightman}, {Nandra}, {Salvato}, {Hsu},
  {Aird}, \& {Rangel}}]{Brightman2014}
{Brightman}, M., {Nandra}, K., {Salvato}, M., {et~al.} 2014, \mnras, 443, 1999

\bibitem[{{Brightman} {et~al.}(2013){Brightman}, {Silverman}, {Mainieri},
  {Ueda}, {Schramm}, {Matsuoka}, {Nagao}, {Steinhardt}, {Kartaltepe},
  {Sanders}, {Treister}, {Shemmer}, {Brandt}, {Brusa}, {Comastri}, {Ho},
  {Lanzuisi}, {Lusso}, {Nandra}, {Salvato}, {Zamorani}, {Akiyama}, {Alexander},
  {Bongiorno}, {Capak}, {Civano}, {Del Moro}, {Doi}, {Elvis}, {Hasinger},
  {Laird}, {Masters}, {Mignoli}, {Ohta}, {Schawinski}, \&
  {Taniguchi}}]{Brightman2013}
{Brightman}, M., {Silverman}, J.~D., {Mainieri}, V., {et~al.} 2013, \mnras,
  433, 2485

\bibitem[{{Buchner} \& {Bauer}(2017)}]{Buchner2017a}
{Buchner}, J. \& {Bauer}, F.~E. 2017, \mnras, 465, 4348

\bibitem[{{Buchner} {et~al.}(2015){Buchner}, {Georgakakis}, {Nandra},
  {Brightman}, {Menzel}, {Liu}, {Hsu}, {Salvato}, {Rangel}, {Aird}, {Merloni},
  \& {Ross}}]{Buchner2015}
{Buchner}, J., {Georgakakis}, A., {Nandra}, K., {et~al.} 2015, \apj, 802, 89

\bibitem[{{Buchner} {et~al.}(2014){Buchner}, {Georgakakis}, {Nandra}, {Hsu},
  {Rangel}, {Brightman}, {Merloni}, {Salvato}, {Donley}, \&
  {Kocevski}}]{Buchner2014}
{Buchner}, J., {Georgakakis}, A., {Nandra}, K., {et~al.} 2014, \aap, 564, A125

\bibitem[{{Buchner} {et~al.}(2017){Buchner}, {Schulze}, \&
  {Bauer}}]{Buchner2017}
{Buchner}, J., {Schulze}, S., \& {Bauer}, F.~E. 2017, \mnras, 464, 4545

\bibitem[{{Burtscher} {et~al.}(2016){Burtscher}, {Davies}, {Graci{\'a}-Carpio},
  {Koss}, {Lin}, {Lutz}, {Nandra}, {Netzer}, {Orban de Xivry}, {Ricci},
  {Rosario}, {Veilleux}, {Contursi}, {Genzel}, {Schnorr-M{\"u}ller},
  {Sternberg}, {Sturm}, \& {Tacconi}}]{Burtscher2016}
{Burtscher}, L., {Davies}, R.~I., {Graci{\'a}-Carpio}, J., {et~al.} 2016, \aap,
  586, A28

\bibitem[{{Dullemond} {et~al.}(2001){Dullemond}, {Dominik}, \&
  {Natta}}]{Dullemond2001}
{Dullemond}, C.~P., {Dominik}, C., \& {Natta}, A. 2001, \apj, 560, 957

\bibitem[{{Dullemond} \& {Monnier}(2010)}]{Dullemond2010}
{Dullemond}, C.~P. \& {Monnier}, J.~D. 2010, \araa, 48, 205

\bibitem[{{Freeman} {et~al.}(2001){Freeman}, {Doe}, \&
  {Siemiginowska}}]{Freeman2001}
{Freeman}, P., {Doe}, S., \& {Siemiginowska}, A. 2001, in Society of
  Photo-Optical Instrumentation Engineers (SPIE) Conference Series, Vol. 4477,
  Society of Photo-Optical Instrumentation Engineers (SPIE) Conference Series,
  ed. J.-L. {Starck} \& F.~D. {Murtagh}, 76--87

\bibitem[{{Fuller} {et~al.}(2016){Fuller}, {Lopez-Rodriguez}, {Packham},
  {Ramos-Almeida}, {Alonso-Herrero}, {Levenson}, {Radomski}, {Ichikawa},
  {Garcia-Bernete}, {Gonzalez-Martin}, {Diaz-Santos}, \&
  {Martinez-Parades}}]{Fuller2016}
{Fuller}, L., {Lopez-Rodriguez}, E., {Packham}, C., {et~al.} 2016, ArXiv
  e-prints

\bibitem[{{Furui} {et~al.}(2016){Furui}, {Fukazawa}, {Odaka}, {Kawaguchi},
  {Ohno}, \& {Hayashi}}]{Furui2016}
{Furui}, S., {Fukazawa}, Y., {Odaka}, H., {et~al.} 2016, \apj, 818, 164

\bibitem[{{Gallagher} {et~al.}(2015){Gallagher}, {Everett}, {Abado}, \&
  {Keating}}]{Gallagher2015}
{Gallagher}, S.~C., {Everett}, J.~E., {Abado}, M.~M., \& {Keating}, S.~K. 2015,
  \mnras, 451, 2991

\bibitem[{{Garc{\'{\i}}a-Burillo} {et~al.}(2016){Garc{\'{\i}}a-Burillo},
  {Combes}, {Ramos Almeida}, {Usero}, {Krips}, {Alonso-Herrero}, {Aalto},
  {Casasola}, {Hunt}, {Mart{\'{\i}}n}, {Viti}, {Colina}, {Costagliola},
  {Eckart}, {Fuente}, {Henkel}, {M{\'a}rquez}, {Neri}, {Schinnerer}, {Tacconi},
  \& {van der Werf}}]{Garcia-Burillo2016}
{Garc{\'{\i}}a-Burillo}, S., {Combes}, F., {Ramos Almeida}, C., {et~al.} 2016,
  \apjl, 823, L12

\bibitem[{{George} \& {Fabian}(1991)}]{pexmonlines}
{George}, I.~M. \& {Fabian}, A.~C. 1991, \mnras, 249, 352

\bibitem[{{Greenhill} {et~al.}(2003){Greenhill}, {Booth}, {Ellingsen},
  {Herrnstein}, {Jauncey}, {McCulloch}, {Moran}, {Norris}, {Reynolds}, \&
  {Tzioumis}}]{Greenhill2003}
{Greenhill}, L.~J., {Booth}, R.~S., {Ellingsen}, S.~P., {et~al.} 2003, \apj,
  590, 162

\bibitem[{{Harrison} {et~al.}(2013){Harrison}, {Craig}, {Christensen},
  {Hailey}, {Zhang}, {Boggs}, {Stern}, {Cook}, {Forster}, {Giommi},
  {Grefenstette}, {Kim}, {Kitaguchi}, {Koglin}, {Madsen}, {Mao}, {Miyasaka},
  {Mori}, {Perri}, {Pivovaroff}, {Puccetti}, {Rana}, {Westergaard}, {Willis},
  {Zoglauer}, {An}, {Bachetti}, {Barri{\`e}re}, {Bellm}, {Bhalerao},
  {Brejnholt}, {Fuerst}, {Liebe}, {Markwardt}, {Nynka}, {Vogel}, {Walton},
  {Wik}, {Alexander}, {Cominsky}, {Hornschemeier}, {Hornstrup}, {Kaspi},
  {Madejski}, {Matt}, {Molendi}, {Smith}, {Tomsick}, {Ajello}, {Ballantyne},
  {Balokovi{\'c}}, {Barret}, {Bauer}, {Blandford}, {Brandt}, {Brenneman},
  {Chiang}, {Chakrabarty}, {Chenevez}, {Comastri}, {Dufour}, {Elvis}, {Fabian},
  {Farrah}, {Fryer}, {Gotthelf}, {Grindlay}, {Helfand}, {Krivonos}, {Meier},
  {Miller}, {Natalucci}, {Ogle}, {Ofek}, {Ptak}, {Reynolds}, {Rigby},
  {Tagliaferri}, {Thorsett}, {Treister}, \& {Urry}}]{HarrisonNuSTAR2013}
{Harrison}, F.~A., {Craig}, W.~W., {Christensen}, F.~E., {et~al.} 2013, \apj,
  770, 103

\bibitem[{{Hasinger} {et~al.}(2005){Hasinger}, {Miyaji}, \&
  {Schmidt}}]{Hasinger2005}
{Hasinger}, G., {Miyaji}, T., \& {Schmidt}, M. 2005, \aap, 441, 417

\bibitem[{{H{\"o}nig} {et~al.}(2012){H{\"o}nig}, {Kishimoto}, {Antonucci},
  {Marconi}, {Prieto}, {Tristram}, \& {Weigelt}}]{Hoenig2012}
{H{\"o}nig}, S.~F., {Kishimoto}, M., {Antonucci}, R., {et~al.} 2012, \apj, 755,
  149

\bibitem[{{Ichikawa} {et~al.}(2016){Ichikawa}, {Ricci}, {Ueda}, {Matsuoka},
  {Toba}, {Kawamuro}, {Trakhtenbrot}, \& {Koss}}]{Ichikawa2016}
{Ichikawa}, K., {Ricci}, C., {Ueda}, Y., {et~al.} 2016, ArXiv e-prints

\bibitem[{{Ikeda} {et~al.}(2009){Ikeda}, {Awaki}, \& {Terashima}}]{Ikeda2009}
{Ikeda}, S., {Awaki}, H., \& {Terashima}, Y. 2009, \apj, 692, 608

\bibitem[{{Kishimoto} {et~al.}(2011){Kishimoto}, {Hoenig}, {Antonucci},
  {Millour}, {Tristram}, \& {Weigelt}}]{Kishimoto2011}
{Kishimoto}, M., {Hoenig}, S.~F., {Antonucci}, R., {et~al.} 2011, ArXiv
  e-prints

\bibitem[{{Krolik} \& {Begelman}(1988)}]{Krolik1988}
{Krolik}, J.~H. \& {Begelman}, M.~C. 1988, \apj, 329, 702

\bibitem[{{Laor} {et~al.}(2006){Laor}, {Barth}, {Ho}, \&
  {Filippenko}}]{Laor2006}
{Laor}, A., {Barth}, A.~J., {Ho}, L.~C., \& {Filippenko}, A.~V. 2006, \apj,
  636, 83

\bibitem[{{Lightman} \& {White}(1988)}]{Lightman1988}
{Lightman}, A.~P. \& {White}, T.~R. 1988, \apj, 335, 57

\bibitem[{{Lira} {et~al.}(2013){Lira}, {Videla}, {Wu}, {Alonso-Herrero},
  {Alexander}, \& {Ward}}]{Lira2013}
{Lira}, P., {Videla}, L., {Wu}, Y., {et~al.} 2013, \apj, 764, 159

\bibitem[{{Liu} \& {Li}(2014)}]{Liu2014}
{Liu}, Y. \& {Li}, X. 2014, \apj, 787, 52

\bibitem[{{Liu} \& {Li}(2015)}]{Liu2015}
{Liu}, Y. \& {Li}, X. 2015, \mnras, 448, L53

\bibitem[{{Magdziarz} \& {Zdziarski}(1995)}]{Magdziarz1995}
{Magdziarz}, P. \& {Zdziarski}, A.~A. 1995, \mnras, 273, 837

\bibitem[{{Marinucci} {et~al.}(2016){Marinucci}, {Bianchi}, {Matt},
  {Alexander}, {Balokovi{\'c}}, {Bauer}, {Brandt}, {Gandhi}, {Guainazzi},
  {Harrison}, {Iwasawa}, {Koss}, {Madsen}, {Nicastro}, {Puccetti}, {Ricci},
  {Stern}, \& {Walton}}]{Marinucci2016}
{Marinucci}, A., {Bianchi}, S., {Matt}, G., {et~al.} 2016, \mnras, 456, L94

\bibitem[{{Marinucci} {et~al.}(2011){Marinucci}, {Bianchi}, {Matt}, {Fabian},
  {Iwasawa}, {Miniutti}, \& {Piconcelli}}]{Marinucci2011}
{Marinucci}, A., {Bianchi}, S., {Matt}, G., {et~al.} 2011, \aap, 526, A36

\bibitem[{{Markowitz} {et~al.}(2014){Markowitz}, {Krumpe}, \&
  {Nikutta}}]{Markowitz2014}
{Markowitz}, A.~G., {Krumpe}, M., \& {Nikutta}, R. 2014, \mnras, 439, 1403

\bibitem[{{Matt}(2000)}]{Matt2000}
{Matt}, G. 2000, \aap, 355, L31

\bibitem[{{Matt} {et~al.}(2000){Matt}, {Fabian}, {Guainazzi}, {Iwasawa},
  {Bassani}, \& {Malaguti}}]{Matt2000a}
{Matt}, G., {Fabian}, A.~C., {Guainazzi}, M., {et~al.} 2000, \mnras, 318, 173

\bibitem[{{McAlary} {et~al.}(1983){McAlary}, {McLaren}, {McGonegal}, \&
  {Maza}}]{McAlary1983}
{McAlary}, C.~W., {McLaren}, R.~A., {McGonegal}, R.~J., \& {Maza}, J. 1983,
  \apjs, 52, 341

\bibitem[{{Murphy} \& {Yaqoob}(2009)}]{MurphyYaqoobMyTorus2009}
{Murphy}, K.~D. \& {Yaqoob}, T. 2009, \mnras, 397, 1549

\bibitem[{{Nandra} \& {George}(1994)}]{Nandra1994a}
{Nandra}, K. \& {George}, I.~M. 1994, \mnras, 267, 974

\bibitem[{{Nandra} {et~al.}(2007){Nandra}, {O'Neill}, {George}, \&
  {Reeves}}]{pexmon}
{Nandra}, K., {O'Neill}, P.~M., {George}, I.~M., \& {Reeves}, J.~N. 2007,
  \mnras, 382, 194

\bibitem[{{Nenkova} {et~al.}(2008{\natexlab{a}}){Nenkova}, {Sirocky},
  {Ivezi{\'c}}, \& {Elitzur}}]{Nenkova2008a}
{Nenkova}, M., {Sirocky}, M.~M., {Ivezi{\'c}}, {\v Z}., \& {Elitzur}, M.
  2008{\natexlab{a}}, \apj, 685, 147

\bibitem[{{Nenkova} {et~al.}(2008{\natexlab{b}}){Nenkova}, {Sirocky},
  {Nikutta}, {Ivezi{\'c}}, \& {Elitzur}}]{Nenkova2008}
{Nenkova}, M., {Sirocky}, M.~M., {Nikutta}, R., {Ivezi{\'c}}, {\v Z}., \&
  {Elitzur}, M. 2008{\natexlab{b}}, \apj, 685, 160

\bibitem[{{Netzer}(2015)}]{Netzer2015}
{Netzer}, H. 2015, \araa, 53, 365

\bibitem[{{Nowak} {et~al.}(2012){Nowak}, {Neilsen}, {Markoff}, {Baganoff},
  {Porquet}, {Grosso}, {Levin}, {Houck}, {Eckart}, {Falcke}, {Ji}, {Miller}, \&
  {Wang}}]{Nowak2012}
{Nowak}, M.~A., {Neilsen}, J., {Markoff}, S.~B., {et~al.} 2012, \apj, 759, 95

\bibitem[{{Paltani} \& {Ricci}(2017)}]{Paltani2017}
{Paltani}, S. \& {Ricci}, C. 2017, \aap, 607, A31

\bibitem[{{Pier} \& {Krolik}(1993)}]{Pier1993}
{Pier}, E.~A. \& {Krolik}, J.~H. 1993, \apj, 418, 673

\bibitem[{{Predehl} \& {Schmitt}(1995)}]{Predehl1995}
{Predehl}, P. \& {Schmitt}, J.~H.~M.~M. 1995, \aap, 293

\bibitem[{{Ramos Almeida} {et~al.}(2009){Ramos Almeida}, {Levenson},
  {Rodr{\'{\i}}guez Espinosa}, {Alonso-Herrero}, {Asensio Ramos}, {Radomski},
  {Packham}, {Fisher}, \& {Telesco}}]{RamosAlmeida2009}
{Ramos Almeida}, C., {Levenson}, N.~A., {Rodr{\'{\i}}guez Espinosa}, J.~M.,
  {et~al.} 2009, \apj, 702, 1127

\bibitem[{{Ricci} {et~al.}(2016){Ricci}, {Bauer}, {Arevalo}, {Boggs}, {Brandt},
  {Christensen}, {Craig}, {Gandhi}, {Hailey}, {Harrison}, {Koss}, {Markwardt},
  {Stern}, {Treister}, \& {Zhang}}]{Ricci2016a}
{Ricci}, C., {Bauer}, F.~E., {Arevalo}, P., {et~al.} 2016, \apj, 820, 5

\bibitem[{{Ricci} {et~al.}(2017){Ricci}, {Trakhtenbrot}, {Koss}, {Ueda}, {Del
  Vecchio}, {Treister}, {Schawinski}, {Paltani}, {Oh}, {Lamperti}, {Berney},
  {Gandhi}, {Ichikawa}, {Bauer}, {Ho}, {Asmus}, {Beckmann}, {Soldi},
  {Balokovic}, {Gehrels}, \& {Markwardt}}]{Ricci2017a}
{Ricci}, C., {Trakhtenbrot}, B., {Koss}, M.~J., {et~al.} 2017, ArXiv e-prints

\bibitem[{{Ricci} {et~al.}(2015){Ricci}, {Ueda}, {Koss}, {Trakhtenbrot},
  {Bauer}, \& {Gandhi}}]{Ricci2016}
{Ricci}, C., {Ueda}, Y., {Koss}, M.~J., {et~al.} 2015, \apjl, 815, L13

\bibitem[{{Risaliti} {et~al.}(2005){Risaliti}, {Elvis}, {Fabbiano}, {Baldi}, \&
  {Zezas}}]{Risaliti2005}
{Risaliti}, G., {Elvis}, M., {Fabbiano}, G., {Baldi}, A., \& {Zezas}, A. 2005,
  \apjl, 623, L93

\bibitem[{{Risaliti} {et~al.}(2007){Risaliti}, {Elvis}, {Fabbiano}, {Baldi},
  {Zezas}, \& {Salvati}}]{Risaliti2007}
{Risaliti}, G., {Elvis}, M., {Fabbiano}, G., {et~al.} 2007, \apjl, 659, L111

\bibitem[{{Risaliti} {et~al.}(2002){Risaliti}, {Elvis}, \&
  {Nicastro}}]{Risaliti2002}
{Risaliti}, G., {Elvis}, M., \& {Nicastro}, F. 2002, \apj, 571, 234

\bibitem[{{Rivers} {et~al.}(2013){Rivers}, {Markowitz}, \&
  {Rothschild}}]{Rivers2013}
{Rivers}, E., {Markowitz}, A., \& {Rothschild}, R. 2013, \apj, 772, 114

\bibitem[{{Rivers} {et~al.}(2015){Rivers}, {Risaliti}, {Walton}, {Harrison},
  {Ar{\'e}valo}, {Baur}, {Boggs}, {Brenneman}, {Brightman}, {Christensen},
  {Craig}, {F{\"u}rst}, {Hailey}, {Hickox}, {Marinucci}, {Reeves}, {Stern}, \&
  {Zhang}}]{Rivers2015}
{Rivers}, E., {Risaliti}, G., {Walton}, D.~J., {et~al.} 2015, \apj, 804, 107

\bibitem[{{Rybicki} \& {Lightman}(1986)}]{Rybicki1986}
{Rybicki}, G.~B. \& {Lightman}, A.~P. 1986, {Radiative Processes in
  Astrophysics}

\bibitem[{{Simm} {et~al.}(2018){Simm}, {Buchner}, {Merloni}, {Nandra}, {Shen},
  {Erben}, {Coil}, {Willmer}, \& {Schneider}}]{Simm2018}
{Simm}, T., {Buchner}, J., {Merloni}, A., {et~al.} 2018, \mnras, 480, 4912

\bibitem[{{Skrutskie} {et~al.}(2006){Skrutskie}, {Cutri}, {Stiening},
  {Weinberg}, {Schneider}, {Carpenter}, {Beichman}, {Capps}, {Chester},
  {Elias}, {Huchra}, {Liebert}, {Lonsdale}, {Monet}, {Price}, {Seitzer},
  {Jarrett}, {Kirkpatrick}, {Gizis}, {Howard}, {Evans}, {Fowler}, {Fullmer},
  {Hurt}, {Light}, {Kopan}, {Marsh}, {McCallon}, {Tam}, {Van Dyk}, \&
  {Wheelock}}]{Skrutskie2006}
{Skrutskie}, M.~F., {Cutri}, R.~M., {Stiening}, R., {et~al.} 2006, \aj, 131,
  1163

\bibitem[{{Treister} {et~al.}(2004){Treister}, {Urry}, {Chatzichristou},
  {Bauer}, {Alexander}, {Koekemoer}, {Van Duyne}, {Brandt}, {Bergeron},
  {Stern}, {Moustakas}, {Chary}, {Conselice}, {Cristiani}, \&
  {Grogin}}]{Treister2004}
{Treister}, E., {Urry}, C.~M., {Chatzichristou}, E., {et~al.} 2004, \apj, 616,
  123

\bibitem[{{Tristram} {et~al.}(2014){Tristram}, {Burtscher}, {Jaffe},
  {Meisenheimer}, {H{\"o}nig}, {Kishimoto}, {Schartmann}, \&
  {Weigelt}}]{Tristram2014}
{Tristram}, K.~R.~W., {Burtscher}, L., {Jaffe}, W., {et~al.} 2014, \aap, 563,
  A82

\bibitem[{{Tristram} \& {Schartmann}(2011)}]{Tristram2011}
{Tristram}, K.~R.~W. \& {Schartmann}, M. 2011, \aap, 531, A99

\bibitem[{{Ueda} {et~al.}(2014){Ueda}, {Akiyama}, {Hasinger}, {Miyaji}, \&
  {Watson}}]{Ueda2014}
{Ueda}, Y., {Akiyama}, M., {Hasinger}, G., {Miyaji}, T., \& {Watson}, M.~G.
  2014, \apj, 786, 104

\bibitem[{{Ueda} {et~al.}(2003){Ueda}, {Akiyama}, {Ohta}, \&
  {Miyaji}}]{Ueda2003}
{Ueda}, Y., {Akiyama}, M., {Ohta}, K., \& {Miyaji}, T. 2003, \apj, 598, 886

\bibitem[{{Verner} {et~al.}(1996){Verner}, {Ferland}, {Korista}, \&
  {Yakovlev}}]{Verner1996}
{Verner}, D.~A., {Ferland}, G.~J., {Korista}, K.~T., \& {Yakovlev}, D.~G. 1996,
  \apj, 465, 487

\end{thebibliography}

\section{Green function computation}

\label{sec:Green-function-computation}We briefly outline how \texttt{XARS}
computes the response spectrum to an input energy as seen from a distant
observer. The user provides a function that describes the geometry,
and a function that specifies how emerging photons are gridded (e.g.
in viewing angle).

Given an input energy, a set of photons is created, typically originating
at the geometry origin and directed at all angles uniformly. Then
these photons are pumped through the geometry as follows.

The distance traveled by the photon is calculated following an exponential
distribution with expectation unity in units of optical depth. The
optical depth is defined as the cross-section to both photo-electric
absorption and Compton scattering. Because the energies are sufficiently
high, electrons appear unbound during scattering, allowing the Klein-Nishina
approximation to within a small error \citep{Liu2014}. With the distance
in units of $\NH$ to travel, the geometry user function is called
with the current position, direction and distance to travel. It computes
the coordinates of the end point. This is achieved by integrating
from the start point in the direction until the $\NH$ value has been
accumulated. In a simple geometry such as a single unit sphere, the
end point is the direction times distance from the current position,
where the distance is related to the ratio of $\NH$ to the radial
column density through the sphere. In more complex geometries (e.g.
many spheres), gaps between objects have to be considered. In any
case, the user function returns both the end point and whether the
photon has escaped to infinity. After this propagation, all photons
still in the medium are interacting with it. Proportional to the relative
photo-electric absorption to Compton scattering cross-section at the
current energy, the photon is randomly either absorbed or Compton
scattered. Compton scattering follows standard formulae \citep[see][]{Rybicki1986,Brightman2011a,Paltani2017}
and assigns a new energy and direction. In the case of absorption,
there is a chance for the photon to be re-emitted as a fluorescent
line. The ratio of the line yield multiplied with abundance to the
total absorption cross-section of that element gives the relative
probability of line emission \citep[we use the energies and line yields specified in][]{Brightman2011a}.
Based on these proportionalities, we distribute the photon to one
of the fluorescence lines, with a new randomly drawn direction. In
either case (Compton scattering, fluorescent emission), the pumping
repeats from the start. If the photon however left the medium, the
photon energy, location and direction are passed to the second user
function to accumulate the Green function response to the current
input energy.

In each pumping step, millions of photons can be processed simultaneously
through the use of \texttt{numpy} arrays, making \texttt{XARS} efficient
even though it is Python-based.

\section{Irradiation of a single blob under various density profiles\label{sec:profiles} }

\begin{figure}
\begin{centering}
\includegraphics[width=0.8\columnwidth]{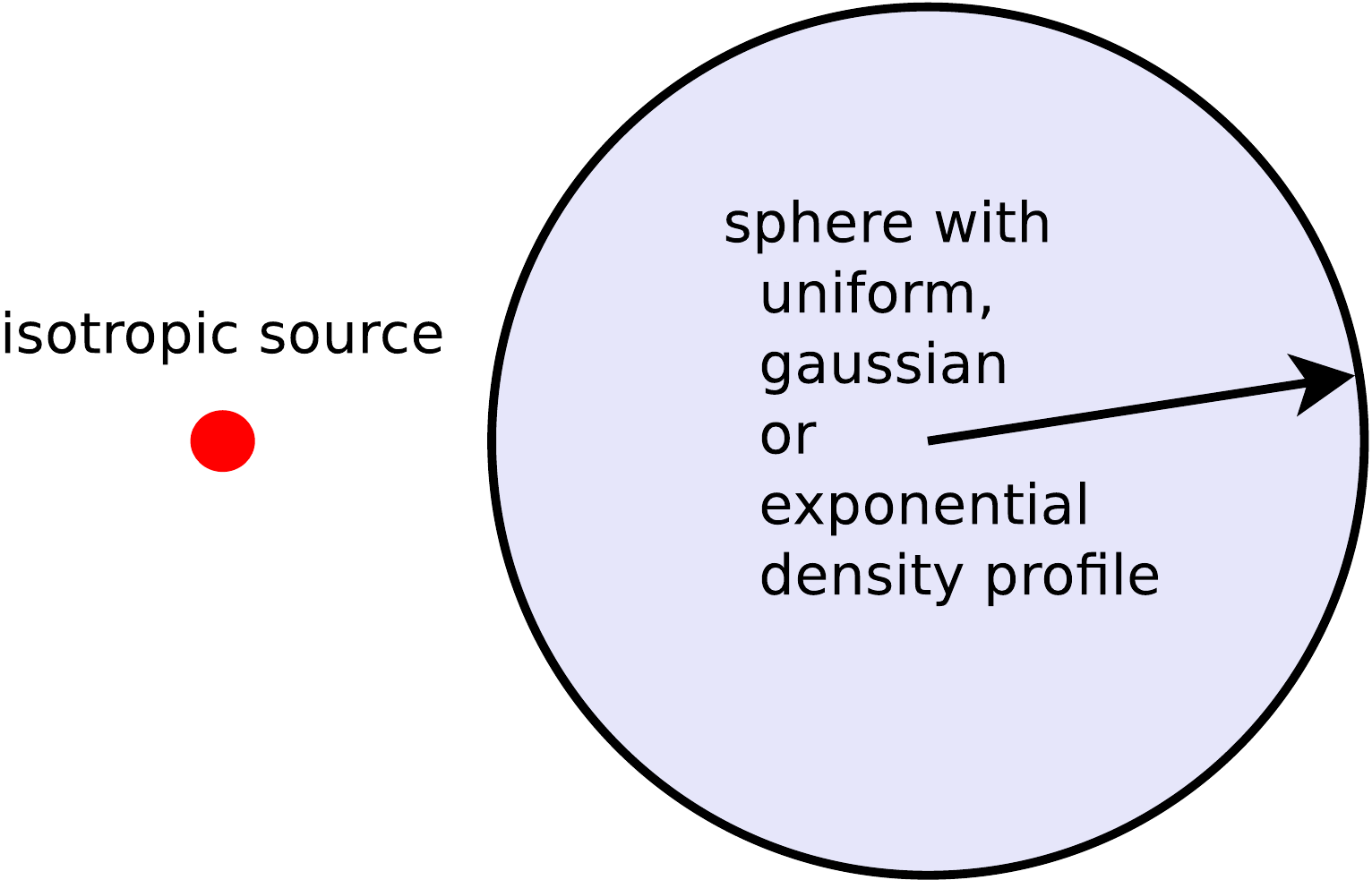}
\par\end{centering}
\caption{\label{fig:Blob-irradiation-geometry}Blob irradiation geometry. A
isotropic source irradiates the blob from a certain distance, and
the spectrum of all reflected photons is assessed in Figure~\ref{fig:Reflection-off-blobs}.}
\end{figure}
\begin{figure}
\begin{centering}
\includegraphics[width=1\columnwidth]{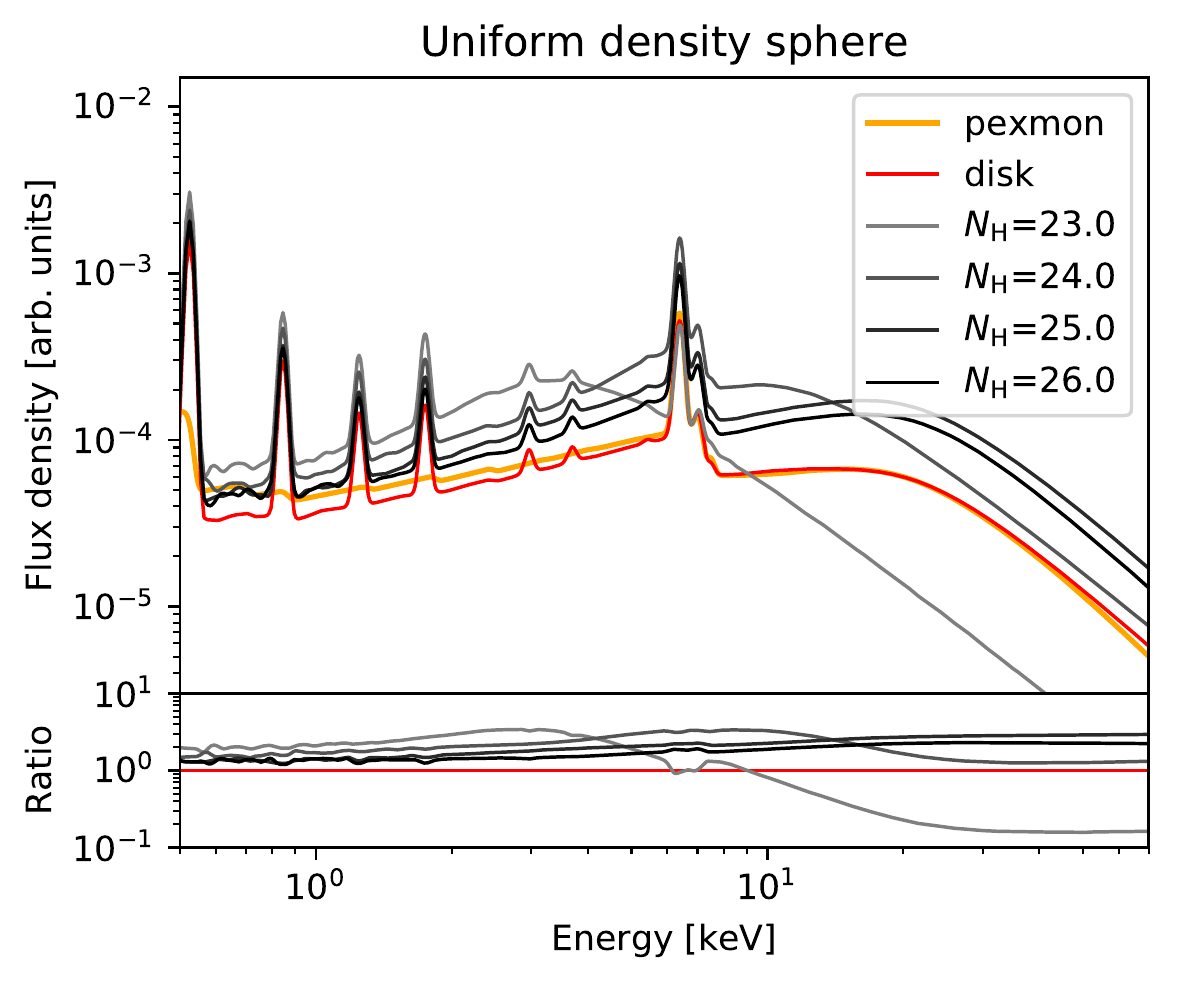}
\par\end{centering}
\begin{centering}
\includegraphics[width=1\columnwidth]{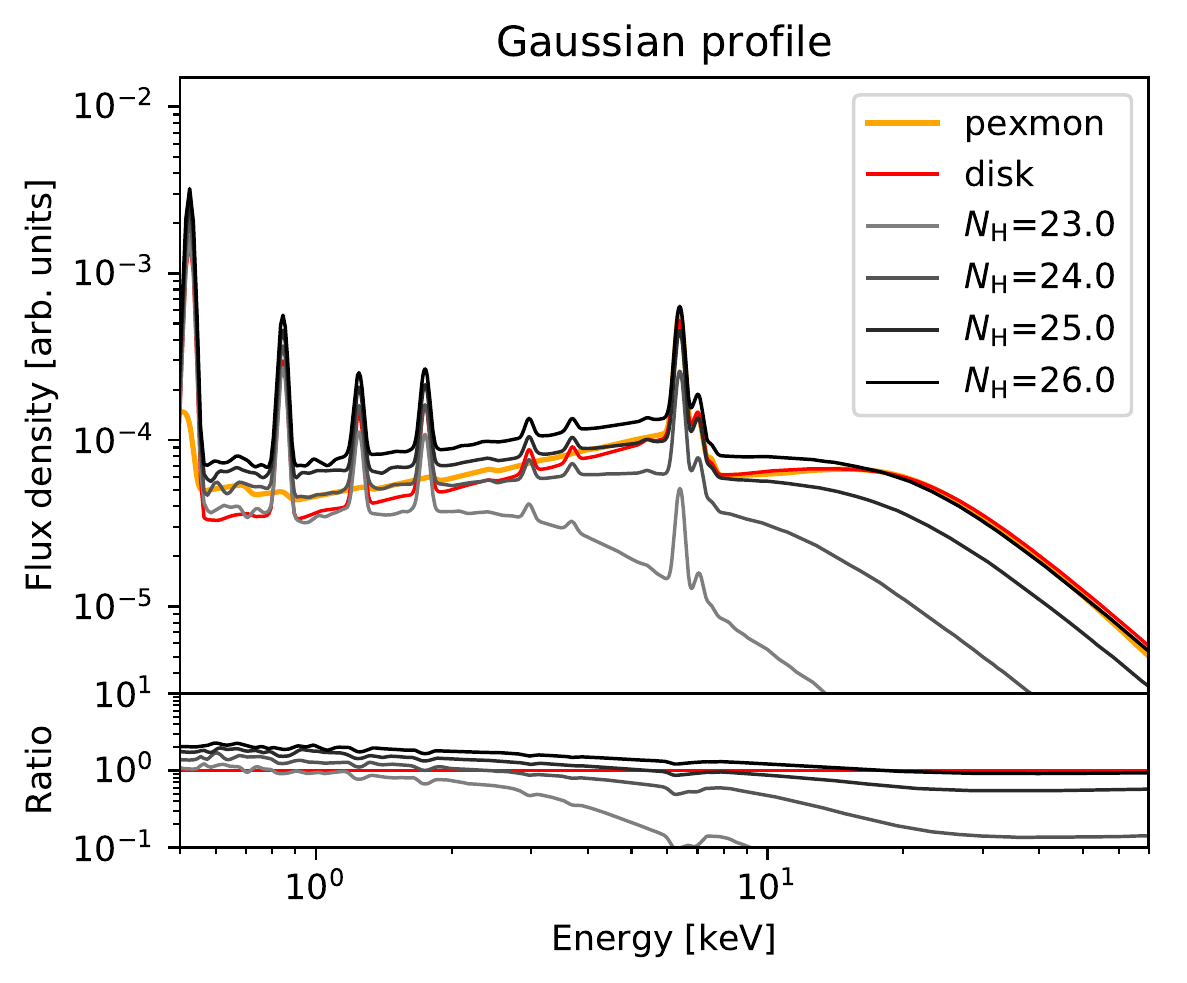}
\par\end{centering}
\begin{centering}
\includegraphics[width=1\columnwidth]{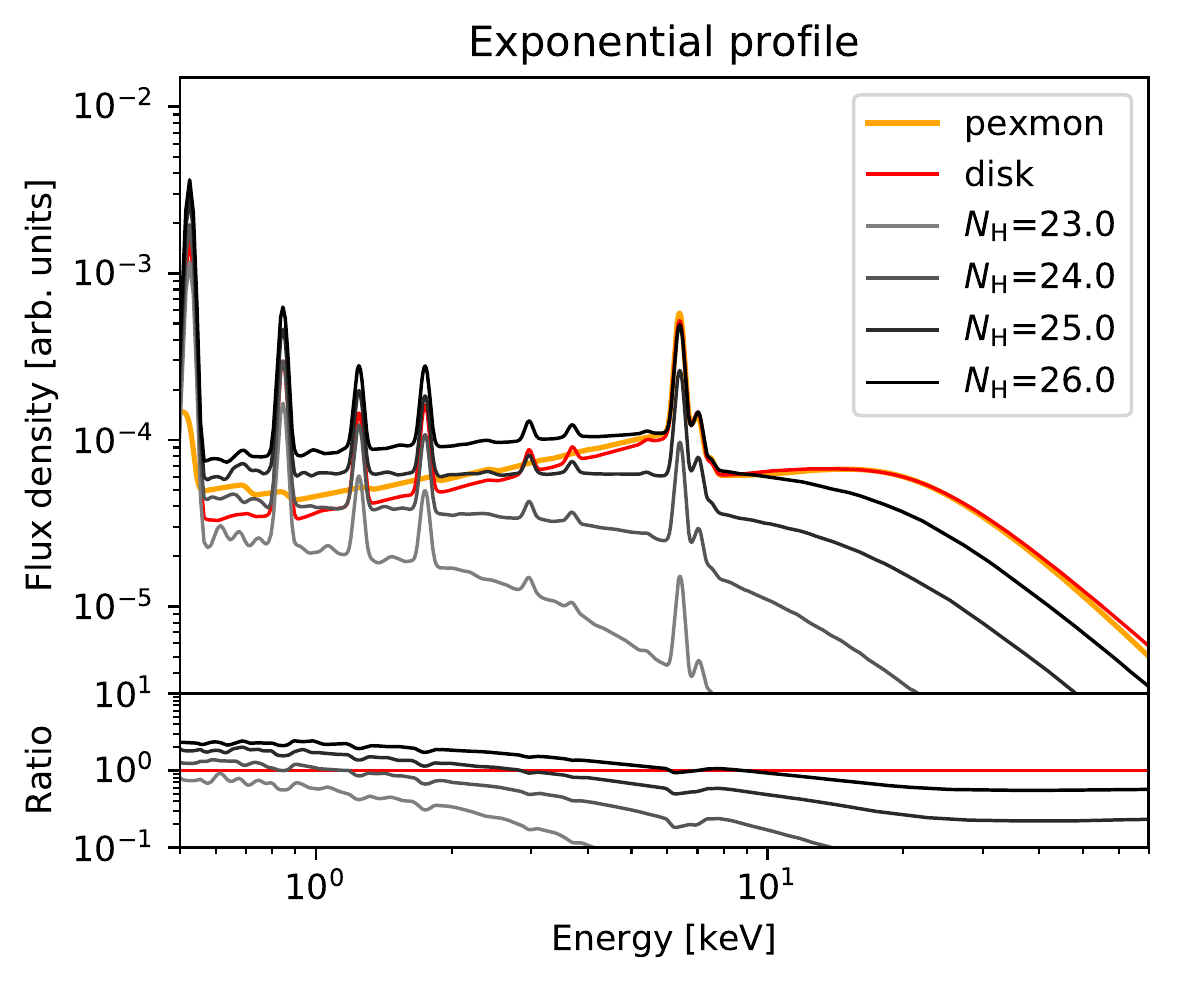}
\par\end{centering}
\caption{\label{fig:Reflection-off-blobs}Reflection off blobs with various
profiles and column densities. For comparison, reflection off a semi-infinite
disk is shown in all panels (orange curve: 50° inclination angle;
red curve: angle-averaged, our code).}
\end{figure}

In this work we have approximated clumps as spheres filled with uniform
density profiles. In this section we determine if more realistic profiles
without sharp cut-offs (e.g. polytropes) would yield different results.
In particular we are interested in the shape of the Compton reflection,
and if varying the density profile could alleviate the need for our
inner ring component.

Our test geometry is simple and shown in Figure~\ref{fig:Blob-irradiation-geometry}:
X-rays from an isotropic source irradiate a single, isolated blob.
We adopt (1) a uniform density profile, (2) a Gaussian profile and
(3) an exponential profile. For the last two we require that the density
at the border is $10^{-6}$ of the centre density. The profiles are
normalised so that maximum column density is $\NH=10^{23,24,25,26}/{\rm cm}^{2}$
(i.e. through the centre). Any reflected/reprocessed X-rays escaping
to infinity, from any angle, are collected and their X-ray spectrum
determined.

The reflection spectra are shown in Figure~\ref{fig:Reflection-off-blobs},
with the panels corresponding to the three profiles. In each panel
we also include reflection off a plane-parallel semi-infinite disk
under a 50° inclination angle (orange line). This is taken from the
\texttt{pexmon} spectrum of \citep{Magdziarz1995,pexmonlines,pexmon},
and comparable to the angle-averaged spectrum produced by our \texttt{XARS}
code of the same geometry (red line). In the top panel, reflection
off high-column density uniform blobs (black lines) resembles the
disk reflection. At lower column densities (grey) the spectrum peaks
at lower energies. The same results were obtained in \citet{Nandra1994a}.
Varying the density profile yields significantly different shapes,
as shown in the other panels of Figure~\ref{fig:Reflection-off-blobs}
for Gaussian and exponential profiles. The Compton hump is much weaker,
and the spectrum flatter towards low energies. This is not just because
these blobs have a lower mass: high-column density results differ
throughout also from results of lower column density uniform spheres.
The location where reflection primarily occurs is at the surface of
uniform blobs, but deeper inside for the other profiles. In the latter,
a photon can pass into the blob further, and after scattering may
escape off the more common lower column density routes. A single scattering
is more likely to put the flight direction of a photon away from the
centre than towards it. In the Gaussian/exponential profiles, this
decreases the effective column density after each scatter, and subsequently
allows high-energy photons to pass through much more easily than in
a uniform density profile.

The spectral shape of \emph{NuSTAR} data studied in this work is similar
to the black lines in the left panel, that is, reflection dominated
with a strong Compton hump. The addition of the torus ring component
emphasises reflection off high-column clouds and surpresses the reflection
off low-density clouds (grey in left panel). As Figure~\ref{fig:Reflection-off-blobs}
shows, introducing Gaussian/exponential profile clumps would worsen
the situation (as their spectra peak at lower energies) and would
under-predict the Compton hump.

\section{Clumpy model distributions \label{sec:Final-model-distributions}}

\begin{figure}
\begin{centering}
\includegraphics[width=1\columnwidth]{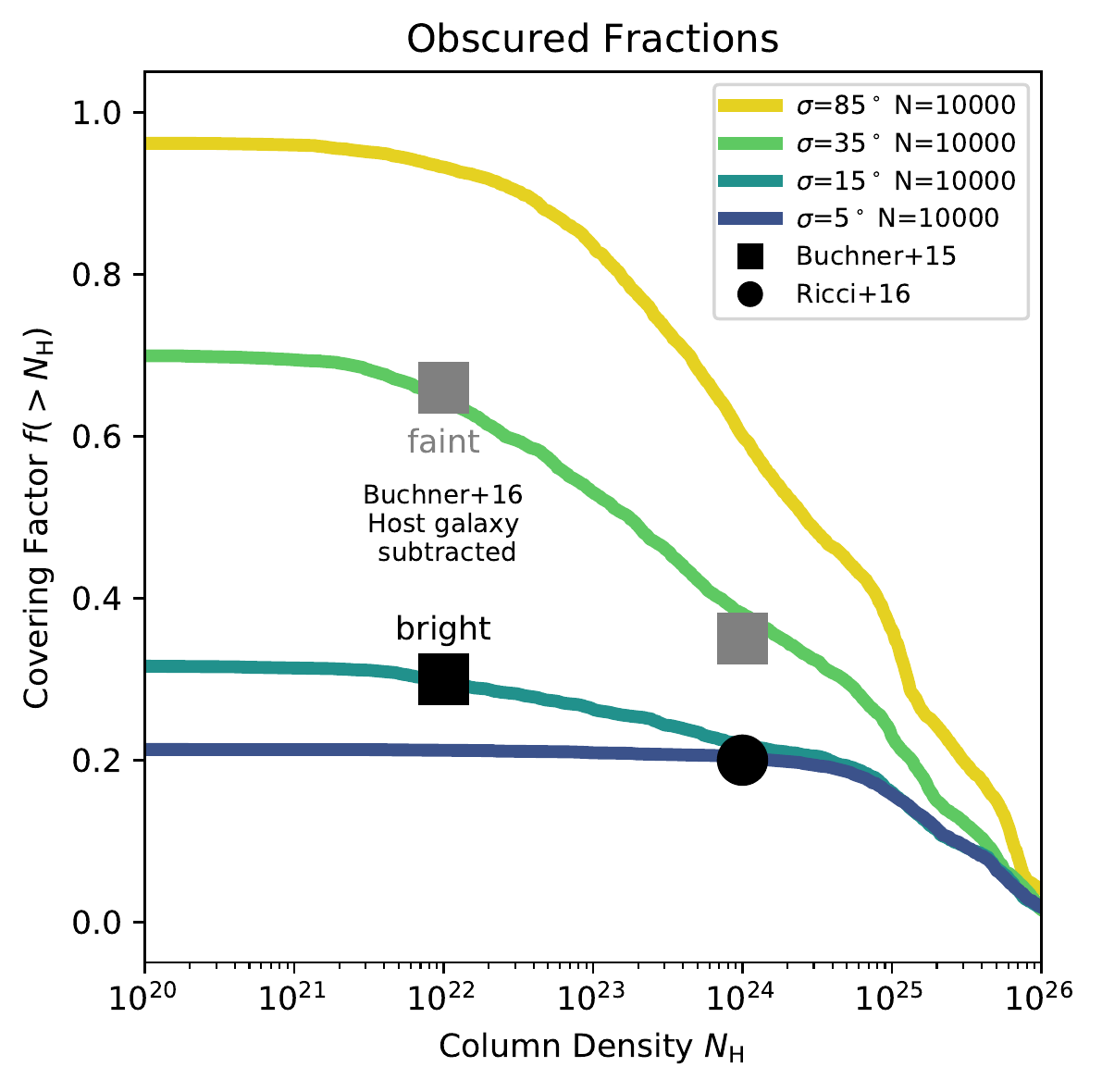}
\par\end{centering}
\caption{\label{fig:Column-density-profile-1}Column density distribution and
associated covering fractions produced by our clumpy models including
the inner ring component. Data points (as in Figure~\ref{fig:Column-density-profile})
represent intrinsic AGN obscured fractions at different luminosities.
Compared to Figure~\ref{fig:Column-density-profile}, the Compton-thick
column densities are more frequent because of the inner Compton-thick
ring.}
\end{figure}

\begin{figure}
\begin{centering}
\includegraphics[width=1\columnwidth]{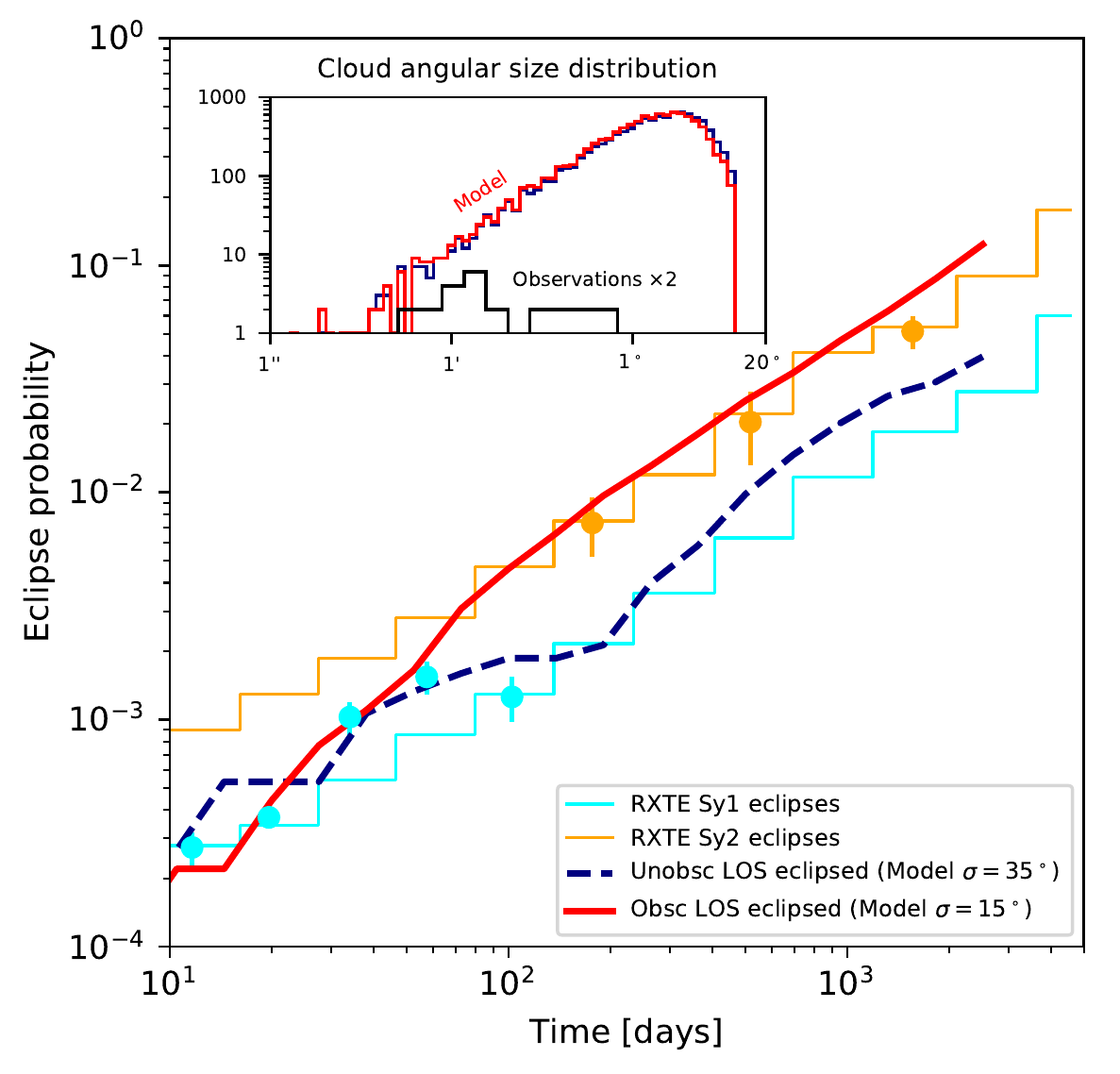}
\par\end{centering}
\caption{\label{fig:eclipses-1}Eclipse rates predicted when including the
inner ring component. Curves as in Figure~\ref{fig:eclipses}, with
orange and cyan error bars and curves representing eclipse measurements
by \citet{Markowitz2014}. Now we show \uxclumpy{} with $\sigma=15\text{°}$
and $\sigma=35\text{°}$ to reproduce Sy1 and Sy2, respectively.}
\end{figure}

Figure~\ref{fig:Column-density-profile-1} presents the column density
distribution of the final model. Similar to Figure~\ref{fig:Column-density-profile},
a wide range of column densities are produced by each geometry. Now,
extremely thick column densities are more common.

Figure~\ref{fig:eclipses-1} presents the frequency of eclipse of
the final model. Similar to Figure~\ref{fig:eclipses}, observed
cloud sizes and eclipse frequencies can be reproduced, albeit with
different model parameters. This is because the inner ring component
shades clouds below its covering factor, requiring higher $\sigma$
values.
\end{document}